\documentclass[aps,twocolumn,prd,showpacs,nofootinbib]{revtex4}
\usepackage{amsmath}
\usepackage{amssymb}
\usepackage{graphicx}
\usepackage{colordvi}

\newcommand{\sss}[1]{{\scriptscriptstyle{#1}}}
\newcommand{\boldmathsymbol}[1]{\ensuremath{\boldsymbol{#1}}}
\newcommand{\order}[1]{\mathcal{O}\!\left(#1\right)}
\newcommand{\Heaviside}[1]{\Theta\!\left(#1\right)}

\newcommand{\besselJ}[1]{J{_{\!#1}}}

\newcommand{\ud}{\mathrm{d}}

\newcommand{\ub}{\mathrm{b}}

\newcommand{\us}{\mathrm{s}}

\newcommand{\ue}{\mathrm{e}}
\newcommand{\ux}{\mathrm{x}}
\newcommand{\uy}{\mathrm{y}}
\newcommand{\ueq}{\mathrm{eq}}
\newcommand{\umat}{\mathrm{mat}}
\newcommand{\urad}{\mathrm{rad}}
\newcommand{\uini}{\mathrm{ini}}
\newcommand{\ucri}{\mathrm{crit}}
\newcommand{\uobs}{\mathrm{obs}}
\newcommand{\uend}{\mathrm{end}}
\newcommand{\ureh}{\mathrm{reh}}
\newcommand{\ur}{\mathrm{r}}
\newcommand{\um}{\mathrm{m}}

\newcommand{\uK}{\mathrm{K}}
\newcommand{\uS}{\mathrm{S}}
\newcommand{\uT}{\mathrm{T}}
\newcommand{\ugw}{\mathrm{gw}}
\newcommand{\uxini}{{\ux\uini}}
\newcommand{\uxend}{{\ux\uend}}
\newcommand{\uyini}{{\uy\uini}}
\newcommand{\uyend}{{\uy\uend}}
\newcommand{\usky}{\mathrm{sky}}
\newcommand{\upix}{\mathrm{pix}}

\newcommand{\calP}{\mathcal{P}}
\newcommand{\calT}{\mathcal{T}}
\newcommand{\calF}{\mathcal{F}}

\newcommand{\calH}{\mathcal{H}}

\newcommand{\Hz}{\mathrm{Hz}}
\newcommand{\Mpc}{\mathrm{Mpc}}

\newcommand{\GeV}{\mathrm{GeV}}

\newcommand{\muK}{\mu\uK}

\newcommand{\bk}{\boldmathsymbol{k}}
\newcommand{\bx}{\boldmathsymbol{x}}

\newcommand{\Mpl}{M_{\mathrm{pl}}}

\newcommand{\rdof}{\mathcal{Q}}
\newcommand{\gstarS}{{g_{\us}}}
\newcommand{\gstarE}{g}
\newcommand{\gstarSzero}{\gstarS_0}
\newcommand{\gstarSreh}{\gstarS_\ureh}
\newcommand{\gstarEzero}{\gstarE_0}
\newcommand{\gstarEreh}{\gstarE_\ureh}
\newcommand{\gs}{{g_*}}

\newcommand{\gsxend}{\gs_\uxend}
\newcommand{\gsyend}{\gs_\uyend}
\newcommand{\gsreh}{\gs_\ureh}

\newcommand{\epsone}{{\epsilon_{1}}}
\newcommand{\epstwo}{{\epsilon_{2}}}
\newcommand{\polartens}{\varepsilon}
\newcommand{\nS}{n_{\sss{\uS}}}
\newcommand{\nT}{n_{\sss{\uT}}}
\newcommand{\calPobs}{\calP^{(\uobs)}}

\newcommand{\Rrad}{R_\urad}

\newcommand{\Rx}{R_\ux}
\newcommand{\Ry}{R_\uy}

\newcommand{\Fy}{F}
\newcommand{\rhoreh}{\rho_\ureh}
\newcommand{\rhoeq}{\rho_\ueq}
\newcommand{\rhoend}{\rho_\uend}
\newcommand{\rhoxini}{\rho_{\uxini}}
\newcommand{\rhoxend}{\rho_{\uxend}}
\newcommand{\rhoyini}{\rho_{\uyini}}
\newcommand{\rhoyend}{\rho_{\uyend}}

\newcommand{\radnow}{{\gamma_0}}
\newcommand{\rhoradnow}{\rho_{\radnow}}
\newcommand{\rhotradnow}{\tilde{\rho}_{\radnow}}

\newcommand{\rhogw}{\rho_\ugw}
\newcommand{\rhocri}{\rho_\ucri}
\newcommand{\azero}{a_0}
\newcommand{\Nend}{N_{\uend}}
\newcommand{\Nreh}{N_\ureh}
\newcommand{\Nstar}{N_*}
\newcommand{\Nk}{\Delta \Nstar}
\newcommand{\Nzero}{N_0}
\newcommand{\wreh}{w_\ureh}
\newcommand{\wrehb}{\overline{w}_\ureh}
\newcommand{\wx}{w_\ux}
\newcommand{\wxb}{\overline{w}_\ux}
\newcommand{\wy}{w_\uy}

\newcommand{\OmegaR}{\Omega_{\ur_0}}
\newcommand{\OmegaM}{\Omega_{\um_0}}

\newcommand{\OmegaGW}{\Omega_\ugw}

\newcommand{\zeq}{z_\ueq}

\newcommand{\zyend}{z_\uyend}
\newcommand{\zxend}{z_\uxend}
\newcommand{\zreh}{z_\ureh}

\newcommand{\aeq}{a_\ueq}
\newcommand{\ak}{a_k}
\newcommand{\ayend}{a_\uyend}
\newcommand{\axend}{a_\uxend}
\newcommand{\areh}{a_\ureh}

\newcommand{\keq}{k_\ueq}
\newcommand{\kyend}{k_\uyend}
\newcommand{\kxend}{k_\uxend}
\newcommand{\kreh}{k_\ureh}

\newcommand{\feq}{f_\ueq}
\newcommand{\fyend}{f_\uyend}
\newcommand{\fxend}{f_\uxend}
\newcommand{\freh}{f_\ureh}
\newcommand{\Treh}{T_\ureh}
\newcommand{\Tyend}{T_\sigma}

\newcommand{\Cov}{\mathrm{Cov}}
\newcommand{\fsky}{f_\usky}
\newcommand{\Npix}{N_\upix}
\newcommand{\tobs}{t_\uobs}
\newcommand{\Sshot}{S_{\mathrm{shot}}}
\newcommand{\Saccel}{S_{\mathrm{accel}}}

\begin{document}

\title{Early Universe Tomography with CMB and Gravitational Waves}

\author{Sachiko Kuroyanagi}
\email{skuro@resceu.s.u-tokyo.ac.jp}
\affiliation{Research Center for the Early Universe, University of Tokyo, Tokyo 113-0033, Japan}

\author{Christophe Ringeval}
\email{christophe.ringeval@uclouvain.be}
\affiliation{Centre for
  Cosmology, Particle Physics and Phenomenology, \\ Institute of
  Mathematics and Physics, Louvain
  University, 2 Chemin du Cyclotron, 1348 Louvain-la-Neuve, Belgium}

\author{Tomo Takahashi}
\email{tomot@cc.saga-u.ac.jp}
\affiliation{Department of Physics, Saga University, Saga 840-8502, Japan}

\date{\today}

\begin{abstract}
  
  We discuss how one can reconstruct the thermal history of the
  Universe by combining cosmic microwave background (CMB) measurements
  and gravitational wave (GW) direct detection experiments. Assuming
  various expansion eras to take place after the inflationary reheating
  and before Big-Bang Nucleosynthesis (BBN), we show how measurements
  of the GW spectrum can be used to break the degeneracies associated
  with CMB data, the latter being sensitive to the total amount of
  cosmic expansion only. In this context, we argue that the expected
  constraints from future CMB and GW experiments can probe a scenario
  in which there exists late-time entropy production in addition to
  the standard reheating. We show that, for some cases, combining data
  from future CMB and GW direct detection experiments allows the
  determination of the reheating temperature, the amount of entropy
  produced and the temperature at which the standard radiation era
  started.

\end{abstract}

\maketitle

\section{Introduction}
\label{sec:intro}

Our understanding of the evolution of the Universe is now becoming
clearer owing to precise cosmological observations such as cosmic
microwave background, large scale structure (LSS), type Ia supernovae
and others. From such observations, we can obtain information about
the current energy budget and the history of the Universe. In
particular, the evolution after the time of Big-Bang Nucleosynthesis
to the present is relatively well understood.  On the other hand, one
can also probe the evolution during inflation since cosmic density
fluctuations, which can be probed by CMB and LSS, are considered to be
initially generated during that epoch.

Compared to the the above mentioned eras, the evolution, or thermal
history of the Universe during the period after inflation to BBN is
relatively unexplored, certainly due to the lack of associated
cosmological observables.  Although, in the standard scenario, the
Universe is considered to be radiation-dominated until BBN (precisely
speaking, until the radiation-matter equality epoch) after the
inflaton reheating, the thermal history can be more complicated.  In
theories beyond the standard model of particle physics such as in
supersymmetric models and string theory, there can exist some scalar
fields (other than the inflaton, as for instance moduli field) that
are long-lived and can dominate the energy density of the
Universe. Their decay may also produce huge amount of entropy thereby
influencing the early universe history.

In the light of these considerations, it would be worth investigating
how one could probe the thermal history during these epochs. In fact,
some authors have investigated this issue by using observations of
CMB~\cite{Martin:2006rs, Ringeval:2007am, Martin:2010kz} and direct
detection of gravitational waves (GW) in Refs.~\cite{Seto:2003kc,
  Nakayama:2008ip, Nakayama:2008wy, Kuroyanagi:2009br}, while using
the combination of both has been pushed forward in
Ref.~\cite{Durrer:2011bi}. From CMB observations, we can probe the
amplitude of the primordial scalar and tensor fluctuations as well as
their scale dependencies around the so-called pivot scale. Notice that
the time at which such a reference scale exited the Hubble radius
during inflation depends on the amount of cosmic expansion from Hubble
exit to the present times, which of course includes all of the
above-mentioned post-inflationary eras. As a result, by measuring the
primordial power spectra in a given inflationary model, we can obtain
information on the amount of the total cosmic expansion, i.e. the
integrated thermal history since the end of inflation.  This can be
also applicable in GW direct detection experiments through
measurements of the amplitude and scale dependence of tensor
fluctuations.  Not only that, direct detection of GWs could be used to
probe the background evolution as the shape of the GW's spectrum is
very sensitive to it. Thus, in the inflationary framework,
detection/non-detection of GWs can give invaluable information on the
thermal history of the Universe thereby allowing a ``tomography'' of
these eras.

In this paper, we investigate this issue by complementing observations
of the CMB and GWs, paying particular attention to the period from the
end of inflation to BBN.  For this purpose, we first recall how one
can constrain the thermal history of the Universe from these
experiments.  Although we have not detected any gravitational waves
yet, CMB observations, such as those from the Wilkinson Microwave
Anisotropy Probe (WMAP)~\cite{Komatsu:2010fb}, are precise enough to
already give some constraints within some inflationary
models~\cite{Martin:2010kz}.  However, in the near future, a direct
detection of GWs could be achieved for some inflationary models that
would allow to combine both CMB and GWs experiments. To see this in an
explicit manner, we investigate the projected constraints on the
thermal history of the Universe from future observations of CMB such
as CMBpol and the PLANCK satellite combined with future direct
detection GW experiments such as BBO \cite{Harry:2006fi}, DECIGO
\cite{ Kawamura:2011zz} and Ultimate-DECIGO \cite{Seto:2001qf}.

The organization of this paper is as follows. In the next section, we
give a brief description on how CMB and GWs can probe or constrain the
thermal history after inflation and justify their complementarity. We
also give the current constraints on the thermal history within the
so-called large field model of inflation coming from CMB using WMAP
data. Then in Section~\ref{sec:const}, we present our forecasts
derived from a Fisher matrix analysis based on the above-mentioned
future CMB and GWs experiments. We conclude in the last section.

\section{CMB and GWs as a probe of the thermal history of the Universe}
\label{sec:CMB_GW}

In this section, we describe how one can probe the thermal history of
the Universe with CMB and GW observations in the context of
inflationary cosmology.

We assume that inflation is the origin of both scalar and tensor
perturbations around the Friedmann--Lema\^{\i}tre--Robertson--Walker
(FLRW) metric
\begin{equation}
\ud s^2 = -a^2 (1 + 2 \Phi) \ud \eta^2 + a^2 \left[(1-2 \Psi)
  \delta_{ij} + h_{ij} \right]\ud x^i \ud x^j\,,
\end{equation}
where $\Psi$ and $\Phi$ are the Bardeen potential and $h_{ij}$ is the
transverse and traceless spin two fluctuations.  If inflation is
driven by a slowly-rolling scalar field $\phi$, the quantum
fluctuations of the field--metric system generate an almost scale
invariant power spectrum for both kinds of perturbation. The observable
quantities are those which are conserved on super-Hubble scales, that
is the comoving curvature for scalar perturbations, which reads in the
longitudinal gauge,
\begin{equation}
\zeta(\eta,\bx) \equiv \Psi(\eta,\bx) + \calH \dfrac{\delta
  \phi(\eta,\bx)}{\phi'}\,,
\end{equation}
where $\calH \equiv a H$ is the conformal Hubble parameter and a prime
denotes derivatives with respect to the conformal time $\eta$. The
tensor modes $h_{ij}$ are themselves gauge invariant and conserved on
super-Hubble scales. It is convenient to decompose them on their two
polarization states $h_\lambda$ in Fourier space as
\begin{equation}
h_{ij}(\eta,\bx) = \sum_{\lambda=+,\times} \int \dfrac{\ud k^3}{(2 \pi)^{3/2}}
  h_\lambda(\eta,\bk) \polartens_{ij}^\lambda \ue^{i \bk \cdot \bx}\,,
\end{equation}
where $\polartens_{ij}^\lambda$ are the polarization tensors
satisfying $\polartens_{ij}^\lambda \polartens^{ij}_{\lambda'} = 2
\delta^{\lambda}_{\lambda'}$.

At first order in the slow-roll formalism, the primordial power
spectrum for the scalars is given by~\cite{Schwarz:2001vv}
\begin{equation}
\label{eq:pzprim}
\begin{aligned}
\calP_\zeta & \equiv \dfrac{k^3}{2 \pi^2} \left|\zeta\right|^2 \simeq
\dfrac{H_*^2}{8 \pi^2 \Mpl^2 \epsone_{*}} \\
& \times \left[1 -2 (C+1) \epsone_* -
  C \epstwo_* - (2 \epsone_* + \epstwo_*) \ln \dfrac{k}{k_*} \right],
\end{aligned}
\end{equation}
while the tensor spectrum (sum of polarization included) reads
\begin{equation}
\label{eq:phprim}
\begin{aligned}
\calP_h &\equiv \dfrac{2 k^3}{\pi^2}\left|h\right|^2 \simeq \dfrac{2
  H_*^2}{\pi^2\Mpl^2} \left[1 -2(C+1) \epsone_* - 2 \epsone_* \ln
  \dfrac{k}{k_*} \right].
\end{aligned}
\end{equation}
In these equations, $\Mpl^2 =1/(8 \pi G)$ stands for the reduced
Planck mass, $C =-2 + \ln 2 + \gamma_E \simeq 0.73$ with $\gamma_E$
being the Euler constant and $\epsilon_i$ are the slow-roll parameters
which are defined as\footnote{
  In the literature, the slow-roll parameters defined using the
  potential for the inflaton $V (\phi)$ are also used, which are given
  by $ \epsilon_V = (1/2) \Mpl^2 \left( V'/ V \right)^2$ and $\eta_V =
  \Mpl^2 (V^{\prime\prime} / V )$ with a prime representing the
  derivative with respect to $\phi$.  The correspondence between $\{
  \epsilon_V, \eta_V \}$ and $ \{ \epsilon_1, \epsilon_2 \}$ at
  leading order in slow-roll parameters are :
\begin{equation}
\epsilon_1 = \epsilon_V, 
\qquad
\epsilon_2 = 2 \epsilon_V -  2 \eta_V.
\end{equation}
With these parameters, the spectral index and the tensor-to-scalar
ratio are respectively given by
\begin{equation}
\nS =  1 - 6 \epsilon_{V*} + 2 \eta_{V*},
\qquad
r = 16 \epsilon_{V*}.
\end{equation}
}
\begin{equation}
\epsilon_1 \equiv  - \frac{ \ud \ln H}{\ud N},
\qquad 
\epsilon_2 \equiv  \frac{ \ud \ln \epsilon_1}{ \ud N},
\end{equation}
where $N \equiv \ln a $ is the number of $e$-folds.
The spectral index defined as $ \nS - 1 = d \ln \mathcal{P}_\zeta / d
\ln k$ is given by
\begin{equation}
\nS = 1 - 2 \epsone_* - \epstwo_*.
\end{equation}
The tensor-to-scalar ratio $r$, which is usually used to quantify the
amplitude of the tensor mode, is given by
\begin{equation}
r  \equiv \frac{\calP_h}{\calP_\zeta } = 16 \epsone_*.
\end{equation}
An asterisk  ``$*$'' indicates that the quantities have to be
evaluated at the  time when the pivot mode
$k_*$ crossed the Hubble radius during inflation, i.e. the solution of
\begin{equation}
\label{eq:pivcross}
k_* = a(\eta_*) H(\eta_*).
\end{equation}
Here, we neglect the running of the spectral index, which will not
affect our results for CMB. However, we note that such truncation can
give large deviation for the power spectrum from the one obtained
exact numerical calculation in some models~\cite{Kuroyanagi:2011iw}.

\subsection{CMB constraints on the post-inflationary universe history}

\subsubsection{Standard scenario}

The power spectrum functional forms of Eqs.~(\ref{eq:pzprim}) and
(\ref{eq:phprim}) are usually compared to the current CMB data to
constrain the slow-roll (Hubble flow) parameters, or equivalently the
spectral index and tensor-to-scalar ratio~\cite{Leach:2000yw,
  Martin:2006rs, Peiris:2006sj, Bean:2007eh, Lorenz:2008je,
  Finelli:2009bs, Komatsu:2010fb, Mortonson:2010er}. This is done by
choosing a pivot scale $k_*$ in the observable range, typically
$k_*=0.05\,\Mpc^{-1}$. However, if one assumes an inflationary model,
there is much more to say. Indeed, the tensor-to-scalar ratio, the
spectral index and all other observable quantities are completely
determined by the inflaton potential $V(\phi)$. As discussed earlier,
they have to be evaluated at the time $\eta_*$, which can be
determined by solving Eq.~(\ref{eq:pivcross}).  In order to obtain
$\eta_\ast$, it is compulsory to make assumptions on the subsequent
thermal history of the universe, i.e. including at least the reheating
and preheating stages. In terms of the number of $e$-folds during
inflation, the physical pivot wavenumber is given by
\begin{equation}
\dfrac{k_*}{a} = \dfrac{k_*}{a_0} (1+z_\uend)\ue^{\Nend-N},
\end{equation}
where $z_\uend$ is the redshift at which inflation ended, after $\Nend$
$e$-folds. As shown in Refs.~\cite{Martin:2006rs, Ringeval:2007am,
  Martin:2010hh, Martin:2010kz}, a convenient way to calculate
$z_\uend$ is to introduce the so-called ``reheating parameter''
\begin{equation}
\label{eq:Rrad}
\Rrad \equiv \dfrac{a_\uend}{a_\ureh} \left( \dfrac{\rhoend}{\rhoreh}
\right)^{1/4}.
\end{equation}
The quantity $\Rrad$ encodes all of our ignorance of the subsequent
thermal evolution after the end of inflation and quantifies any
deviations from a pure radiation era. In fact, by assuming
instantaneous transition between the inflationary epoch to inflaton
oscillating era and inflaton oscillating to radiation-dominated eras,
one has
\begin{equation}
  1+z_\uend = \dfrac{1}{\Rrad} \left( \dfrac{\rhoend}{\rhotradnow} \right)^{1/4},
\end{equation}
where $\rhoend$ is the energy density of the universe at the end of
inflation and $\rhotradnow$ is the energy density of radiation today,
eventually rescaled by any change in the number of gravitating
relativistic degrees of freedom. In terms of the cosmological
parameters today,
\begin{equation}
\label{eq:rhotilde}
\rhotradnow = \rdof_\ureh \rhoradnow  = 3 \rdof_\ureh \, \dfrac{H_0^2}{\Mpl^2}
\OmegaR\, ,
\end{equation}
where we have defined
\begin{equation}
\rdof_\ureh \equiv \dfrac{\gstarEreh}{\gstarEzero}
\left(\dfrac{\gstarSzero}{\gstarSreh} \right)^{4/3}.
\label{eq:rdof}
\end{equation}
Here, $\gstarS$ and $\gstarE$ respectively denotes the number of
entropic and energetic relativistic degrees of freedom at the epoch of
interest.  $H_0$ and $\OmegaR$ are the Hubble parameter and radiation
density parameter today.

As shown in Ref.~\cite{Martin:2010hh}, using energy conservation,
Eq.~(\ref{eq:Rrad}) can be recast into two other strictly equivalent
forms
\begin{equation}
\label{eq:Rrad2}
\begin{aligned}
\ln \Rrad & = \dfrac{1}{4} (\Nreh - \Nend) (3 \wrehb - 1) \\
& = \dfrac{1 -
  3 \wrehb}{12(1+ \wrehb)} \ln\left(\dfrac{\rhoreh}{\rhoend} \right),
\end{aligned}
\end{equation}
with $\wrehb$ standing for the \emph{mean} equation of state parameter
during the inflaton oscillating era. Using $\Rrad$,
Eq.~(\ref{eq:pivcross}) is solved for the $e$-fold time $\Nk \equiv
\Nstar - \Nend$,  verifying~\cite{Martin:2010hh}
\begin{equation}
\label{eq:pivsol}
\begin{aligned}
\Nk = -\ln \Rrad  + \Nzero
& - \frac14 \ln\left(\frac{H_*^2}{\Mpl^2\epsone_{*}} \right)
 \\ & + 
\frac{1}{4}\ln\left(\frac{3}{\epsone_{*}}
\frac{V_\uend}{V_*}\frac{3-\epsone_{*}}
{3-\epsone_{\uend}}\right),
\end{aligned}
\end{equation}
where the constant $\Nzero$ stands for
\begin{equation}
  \Nzero \equiv \ln \left( \dfrac{k_*/a_0}{\rhotradnow^{1/4}} \right).
\end{equation}
Let us emphasize that the right hand side of Eq.~(\ref{eq:pivsol})
usually depends on $\Nk$ itself, but in a completely algebraic way
once the model, i.e. $V(\phi)$, is specified. It can also be
simplified further by using
$\epsone_* \ll 1$, $\epsone_\uend=1$ and Eq.~(\ref{eq:pzprim}) as, 
\begin{equation}
\Nk = -\ln \Rrad + \Nzero - \dfrac{1}{4} \ln (8 \pi^2 \calP_*) -
\dfrac{1}{4} \ln \dfrac{r}{72} + \dfrac{1}{4} \ln \dfrac{V_\uend}{V_*}\,.
\end{equation}

\begin{figure}
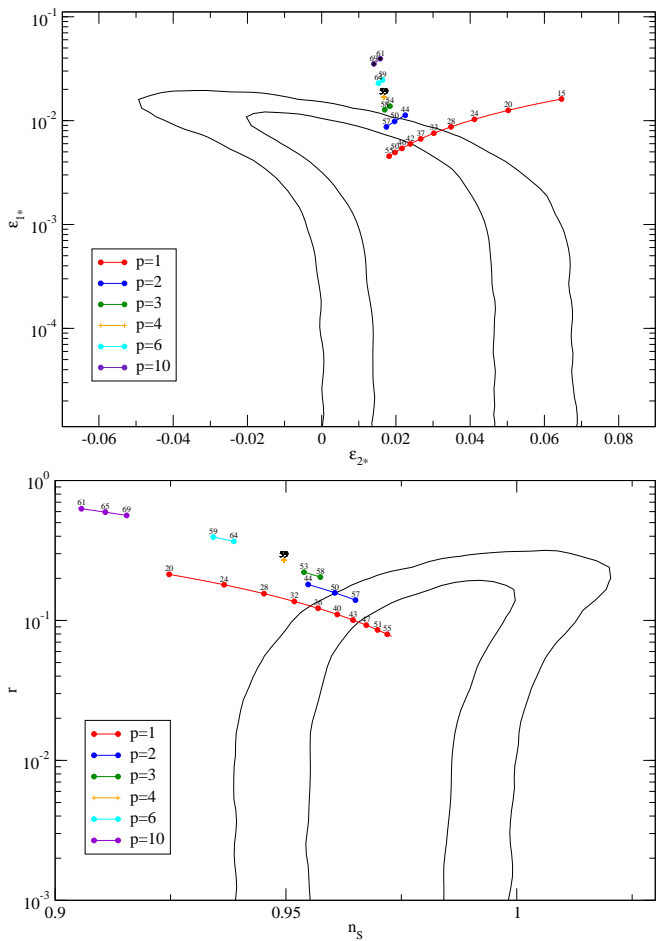

\begin{center}
\includegraphics[width=\linewidth]{lfreh_wmap7.eps}
\includegraphics[width=\linewidth]{lfreh_nsr_wmap7.eps}
\caption{WMAP7 constraints in the plane $(\epsone_*,\epstwo_*)$ (top)
  and $(\nS,r)$ (bottom) compared with the large field model
  predictions obtained by solving Eq.~(\ref{eq:pivsol}) for various
  monomial potentials $V(\phi) \propto \phi^p$. The annotated values
  are those of $|\Nk|$ and they range for a reheating occurring as low
  as BBN to an instantaneous reheating after inflation.}
\label{fig:lfp}
\end{center}
\end{figure}

As an example, we have plotted in Fig.~\ref{fig:lfp} the predicted
values for $\epsone_*$, $\epstwo_*$, as well as the spectral index
$\nS$ and tensor-to-scalar ratio $r$ for the chaotic inflation models
with $V(\phi) \propto \phi^p$.  We show the cases of $p=1,2,3,4,6$ and
$10$. For each $p$, there is a range of possible $\Nk$ since the
reheating should occur from the end of inflation to the BBN epoch,
whose values are indicated in the figure at both end points and for
each case of $p$.  Along with the theoretical predictions, we also
show 1$\sigma$ and 2$\sigma$ confidence intervals associated with the
WMAP7 data~\cite{Larson:2010gs,Jarosik:2010iu,Martin:2010kz} and
Hubble Space Telescope (HST) data~\cite{Riess:2009pu}. From the
figure, one can easily see that inflation models with $p \ge 3$ are
excluded by current data. Even for the cases with $p=2$ and $1$, there
is a lower bound on $|\Nk|$ to be consistent with WMAP7, which can be
translated into the constraints on the thermal history of the
Universe.

\subsubsection{Non-standard scenarios}

Up to here, we have assumed the ``standard" scenario in which once the
reheating from the inflaton is completed, the Universe becomes
radiation dominated until the radiation-matter equality $\zeq \sim 10^4$.  However, this standard scenario could be modified. For
instance, if one assumes that, inserted after the reheating era there
is a phase of evolution dominated by a gravitating source $X$,
characterized by an equation of state parameter $\wx$. As shown in
Ref.~\cite{Martin:2010hh}, one can define a parameter $\Rx$ exactly as
in Eq.~(\ref{eq:Rrad}) by
\begin{equation}
\label{eq:Rx}
\Rx \equiv \dfrac{a_{\uxini}}{a_{\uxend}}
\left(\dfrac{\rhoxini}{\rhoxend} \right)^{1/4},
\end{equation}
for which it is immediate to verify that Eq.~(\ref{eq:Rrad2}) also
applies using the mean value $\wxb$ and by the replacement ``$\uend
\rightarrow \uxini$'', ``$\ureh \rightarrow \uxend$''. From this
definition, one can check that all equations are unchanged, and in
particular Eq.~(\ref{eq:pivsol}), by replacing $\Rrad$ with $\Rrad
\Rx$. Assuming another $Y$-era to take place just after the $X$-era
and before the radiation-dominated era, we would reach exactly the
same conclusions by replacing $\Rrad \Rx$ with $\Rrad \Rx \Ry$. In
other words, CMB can only constrain the overall thermal
history and only feels those parameters, $\Rrad \Rx \Ry \dots$,
multiplied. Let us also notice that the correction coefficient
entering Eq.~(\ref{eq:rhotilde}) is now given by $\rdof_\uyend$
instead of $\rdof_\ureh$.

\begin{figure*}
\begin{center}
\includegraphics[width=0.66\linewidth]{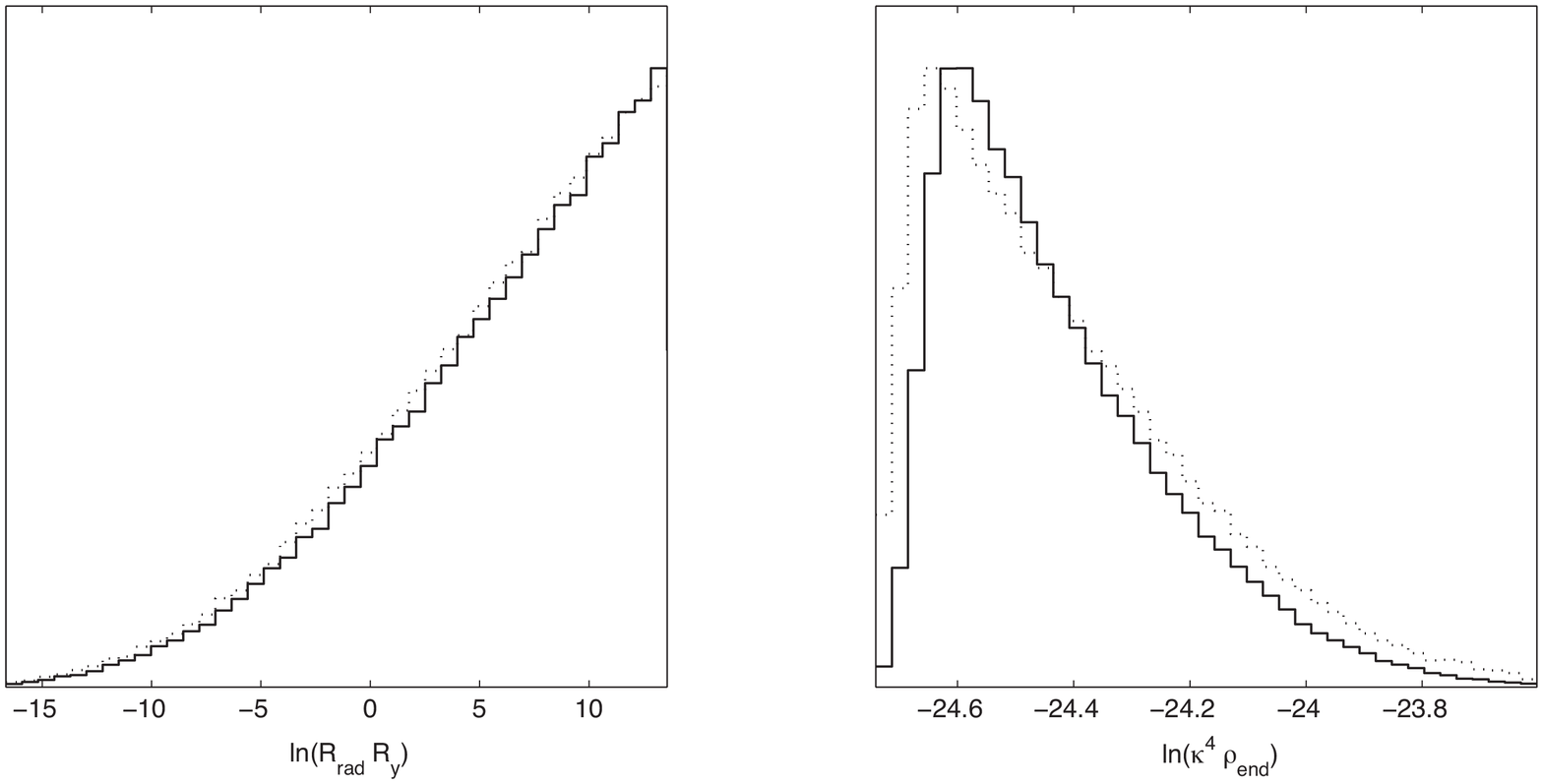}
\includegraphics[width=0.33\linewidth,clip=no]{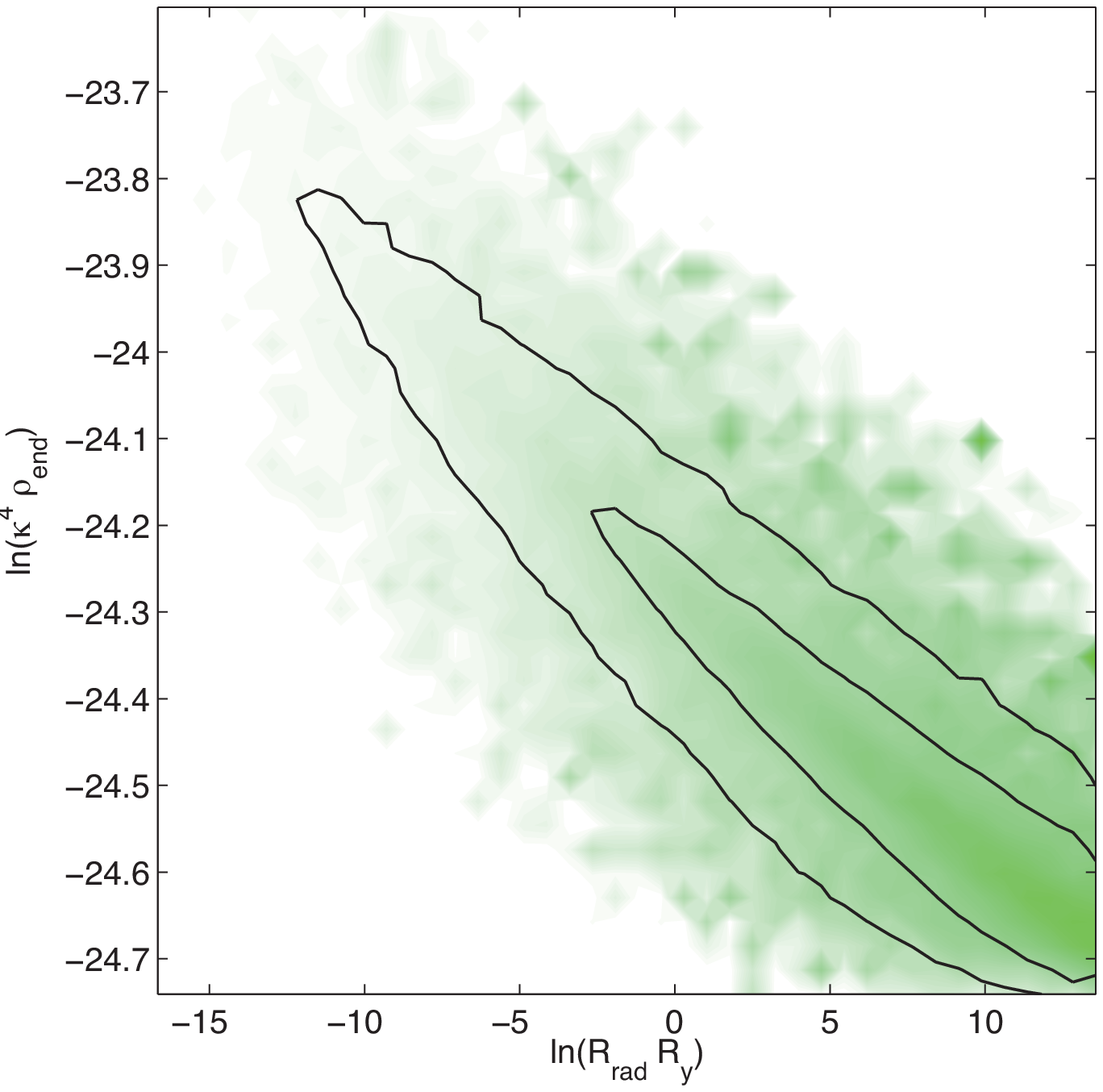}
\caption{Marginalized posterior probability distributions (solid line)
  and mean likelihood (dotted line) from WMAP7 and HST data for
  massive inflation with $p=2$. The right figure shows the one- and
  two-sigma confidence intervals of the two-dimensional marginalized
  posterior in the plane $(\Rrad \Ry,\kappa^4 \rhoend)$.}
\label{fig:1D_wmap_p2}
\end{center}
\end{figure*}

\begin{figure*}
\begin{center}
\includegraphics[width=\linewidth]{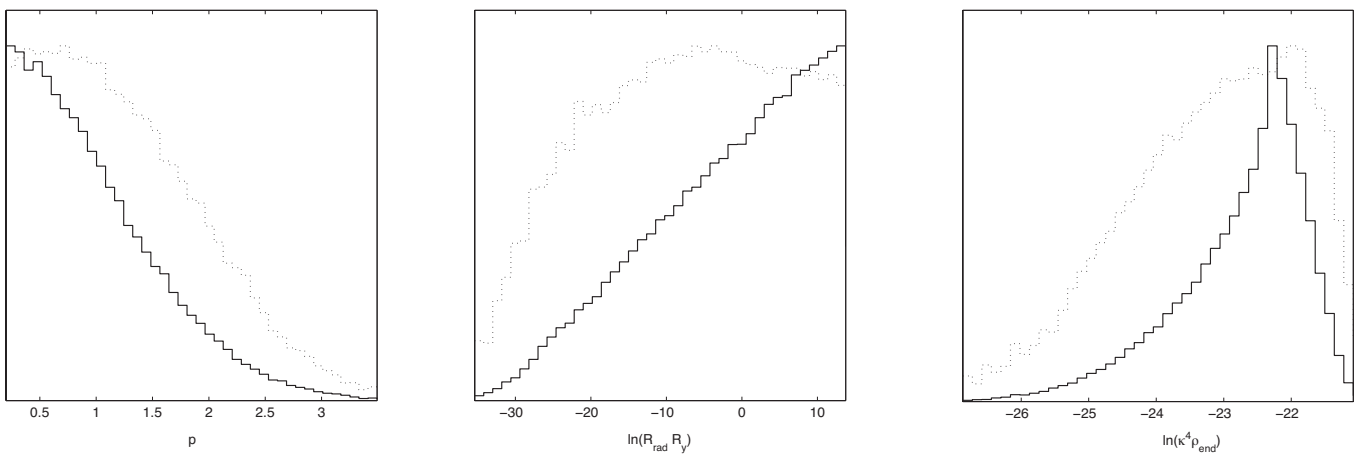}
\caption{Marginalized posterior probability distributions (solid line)
  and mean likelihood (dotted line) from WMAP7 and HST data for the
  large field potential with a power law exponent $p$ being varied in
  $0.2 < p < 5$. }
\label{fig:1D_wmap_pvar}
\end{center}
\end{figure*}

\begin{figure*}
\begin{center}
\includegraphics[width=\linewidth]{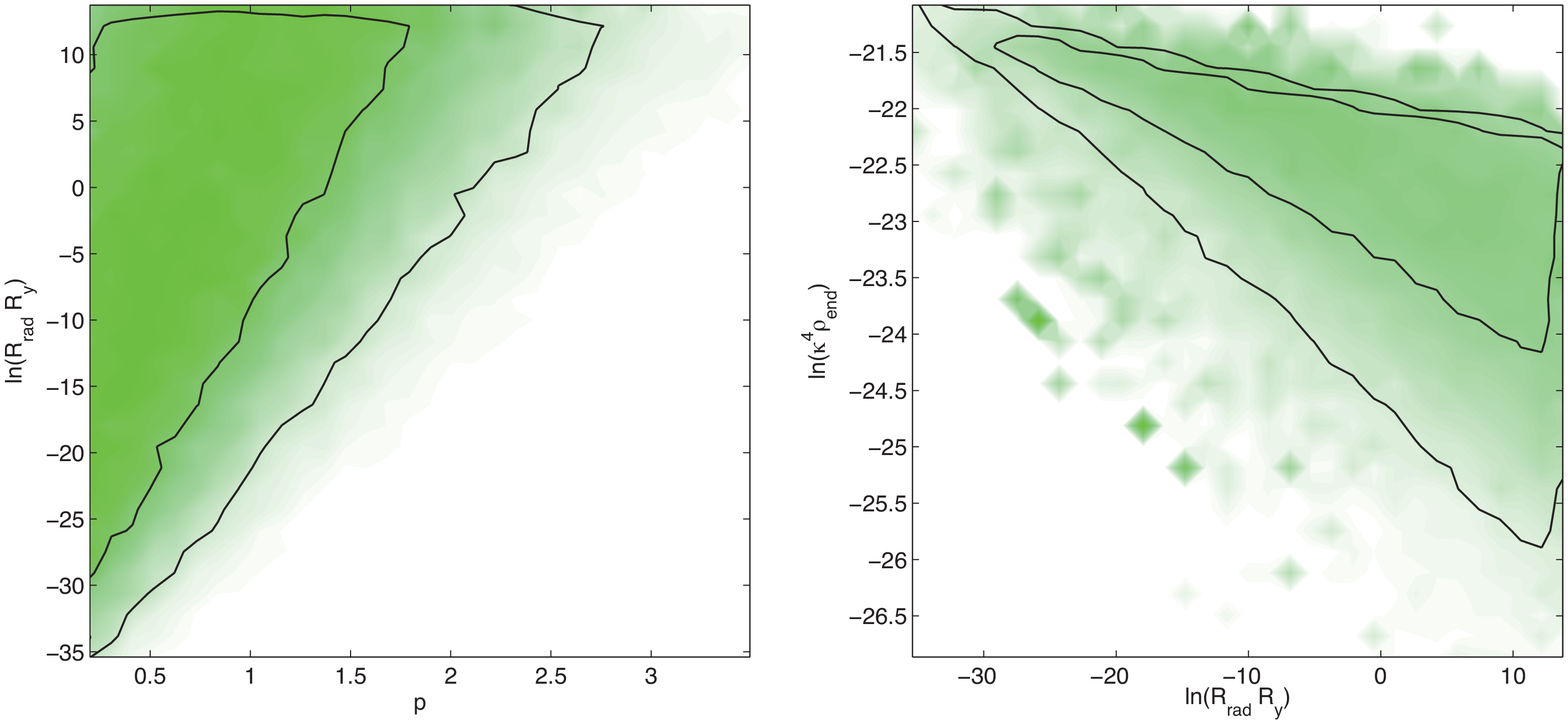}
\caption{Two-dimensional posteriors with 1$\sigma$ and 2$\sigma$
  confidence regions from WMAP7 and HST data in the $(p,\Rrad \Ry)$
  plane (left) and in the $(\Rrad \Ry,\kappa^4 \rhoend)$ plane
  (right).}
\label{fig:2D_wmap_pvar}
\end{center}
\end{figure*}

As a well motivated example, and the one we will be discussing in
Sec.~\ref{sec:const}, such a situation is typical of scenarios in
which a late-decaying massive scalar field, denoted as $\sigma$
hereafter, produce a large amount of entropy well after the inflaton
reheating. In that case, the $X$-era is a short radiation-dominated
era standing just after inflaton reheating and before the $\sigma$
domination. For such a scenario, one has $\Rx=1$ whereas, the $Y$-era
would precisely correspond to the field domination era having $\wy=0$
such that $\Ry$ can only take negative values (quadratic
potential). In Fig.~\ref{fig:1D_wmap_p2} to
Fig.~\ref{fig:2D_wmap_pvar} we have represented the WMAP7 constraints
on the combination $\Rrad \Rx \Ry$ (for $\Rx=1$) for various large
field models, either massive as the scenario we are interested in, or
for any values of $p$, the power law exponent of the inflaton
potential. The method we have used is the same as in
Ref.~\cite{Martin:2010kz} and we do not repeat the details here. As
already shown in Fig.~\ref{fig:lfp}, large values of $|\Nk|$, which
corresponds to $\Rrad \Ry < 0$, give bigger $r$ and more red-tilted
spectral index $\nS$. Hence smaller values of $\Rrad \Ry$ are
disfavored, which means that the current observations already give
some constraints on the thermal history of the Universe. The
posteriors for $\rhoend$ are also depicted and since it is essentially
determined by the amplitude of the curvature perturbation, it is well
bounded~\cite{Martin:2010kz}.

In some literature, when one considers a late-time entropy production
scenario, the parameter $\Fy$ is adopted to quantify the amount of
entropy production, instead of $\Ry$, and is defined as
\begin{equation}
\label{eq:Fy}
\Fy \equiv \frac{s_{\uyend} a_{\uyend}^3}{s_{\uyini} a_{\uyini}^3}\,.
\end{equation}
The subscript ``yini'' and ``yend'' indicate that the quantities are
the ones evaluated at the time when $Y$-era starts and ends. Contrary
to the definition of the $\Rrad$ and $\Rx$ parameters which only
require that total energy density is conserved, the definition of
$\Fy$ is convenient if thermalization is achieved. In our scenario, at
the beginning of the $Y$-era, and also just after its end, the
Universe is assumed to be radiation dominated, and if thermalized, the
entropy density is dominated by relativistic species. In that
situation, it is straightforward to show that
\begin{equation}
\Ry = \Fy^{-1/3} \left(\dfrac{\rdof_\uyini}{\rdof_\uyend}\right)^{1/4}\,,
\end{equation}
where $\rdof$ is defined as in Eq.~(\ref{eq:rdof}) for the epochs of
interest.

\subsection{Stochastic gravitational waves background}

As discussed in the previous section, CMB can constrain the amount of
entropy production but this will remain completely degenerated with
reheating from the inflaton as the only quantities appearing in the
determination of $\Nk$ is the product of $R$ parameters such as $\Rrad
\Rx \Ry$. As we show below, this is not the same for stochastic
gravitational waves of inflationary origin: they feel these parameters
in a different way which can be used to break the degeneracies,
thereby performing the tomography of the history of the Universe. In
particular, direct detection experiments as BBO and DECIGO, which
probe the frequency range $f \sim \order{1}\,\Hz$, would give new and
complementary information with respect to CMB.

In order to discuss the amplitude of stochastic GW, one usually uses
the spectrum of the energy density of GW normalized by the critical
energy density $\rhocri$. From the pseudo stress-tensor, assuming a
stochastic background in which spatial and time averages are
identical, one gets~\cite{Maggiore:1999vm,Kuroyanagi:2010mm}
\begin{equation}
\label{eq:omegaGW}
\begin{aligned}
\OmegaGW \equiv \frac{1}{\rhocri} \frac{\ud \rhogw}{\ud \ln k} & =
\frac{1}{12} \left( \frac{k}{aH} \right)^2 \frac{k^3}{\pi^2}
\sum_\lambda | h_\lambda |^2 \\
& = \frac{1}{12} \left( \frac{k}{aH} \right)^2 \calPobs_h(k).
\end{aligned}
\end{equation}
The observed power spectrum $\calPobs_h$ at the time of interest
(e.g. today) is given by Eq.~(\ref{eq:phprim}) times a transfer
function encoding the evolution of the sub-Hubble modes
\begin{equation}
\calPobs_h(k) = \calP_h(k) \calT^2(k).
\end{equation}

The transfer function $\calT$ can be evaluated analytically since,
after inflation, the tensor perturbations are decoupled from other
sources and their equation of evolution in Fourier space is given by
\begin{equation}
\label{eq:tensevol}
h'' + 2 \calH h' + k^2 h =0\,,
\end{equation}
where for simplicity the polarization index has been dropped. Assuming for the
moment that the background is dominated by a fluid having a constant
equation of state parameter $w$, the solution reads
\begin{equation}
h(\eta,\bk) \propto (k \eta)^{(3 w-3)/(6 w + 2)}
\besselJ{\frac{3-3w}{6w+2}}(k \eta).
\end{equation}
For super-Hubble modes ($k\eta \ll 1$) we recover that $h(\eta,\bk)$
is constant while for sub-Hubble wavenumbers one has
\begin{equation}
|h(\eta,\bk)|^2 \underset{k\eta \gg1}{\propto} a^{-2}\,.
\end{equation}
Since $h$ stays
constant on super-Hubble scales, the transfer function is determined
only by the era at which a given mode $k$ reenters the Hubble
radius. Every mode $k$ will then be damped till the present time
by a factor $(\ak/a_0)^2$ where $\ak$ is solution of
\begin{equation}
\label{eq:reenter}
k = \ak H(\ak)\,,
\end{equation}
and the scale factor today will be taken as unity $a_0=1$.

\subsubsection{Radiation and matter eras}

We start with the case for which the mode reenters during radiation
era. Assuming an instantaneous transition, the result can be
approximated by squaring Eq.~(\ref{eq:reenter}) and using the first
Friedmann--Lema\^{\i}tre equation, $H^2 = \rho/(3 \Mpl^2)$. One gets
\begin{equation}
\label{eq:ak2}
\ak^2 = \dfrac{3 \Mpl^2 k^2}{\rho(\ak)} = \dfrac{3 \Mpl^2 k^2}{\rhoeq
  \aeq^4} \ak^4\,,
\end{equation}
where $\rho \propto a^{-4}$ is used. The solution reads
\begin{equation}
\label{eq:akrad}
\ak = \aeq \dfrac{\keq}{k} = \dfrac{1}{1+\zeq} \dfrac{\keq}{k}\,,
\end{equation}
with
\begin{equation}
\label{eq:keq}
\keq \equiv \dfrac{1}{\sqrt{3} \Mpl} \dfrac{\sqrt{\rhoeq}}{1 + \zeq}
\simeq H_0 \dfrac{\OmegaM}{\sqrt{\OmegaR}}\,.
\end{equation}
In the rightmost side, $\OmegaM$ and $\OmegaR$ are the matter and
radiation density parameters today and $\rhoeq \simeq 2
\rhoradnow/\aeq^4$ is the energy density at equality. For all modes
entering the Hubble radius in the radiation era, Eq.~(\ref{eq:akrad})
immediately gives the transfer function for $k>\keq$
\cite{Turner:1993vb}:
\begin{equation}
\label{eq:trad}
\calT_\urad(k) \simeq H_0^2 \OmegaM \dfrac{\keq}{k} \simeq
\dfrac{H_0 \sqrt{\OmegaR}}{k},
\end{equation}

Similarly, starting from Eq.~(\ref{eq:reenter}), the matter era
solution reads
\begin{equation}
\label{eq:tmat}
\calT_\umat(k) \simeq \dfrac{H_0^2 \OmegaM}{k^2} \Heaviside{k < \keq}\,,
\end{equation} 
where $\Heaviside{x}$ is the Heaviside step function.  In fact,
considering a mixture of radiation and matter, one would find
\begin{equation}
\label{eq:tradmat}
\calT_\ueq(k) \simeq \dfrac{H_0^2 \OmegaM}{2 k^2} \left[1 + \sqrt{1
  + 4 \left(\dfrac{k}{\keq}\right)^2} \right],
\end{equation}
which gives back Eqs.~(\ref{eq:trad}) and (\ref{eq:tmat}) in the
appropriate limits. Although not specified, the above transfer
functions are unity for super-Hubble modes,
i.e. $k<H_0^{-1}$. Combined with Eq.~(\ref{eq:omegaGW}) we recover the
well-known result~\cite{Kuroyanagi:2010mm} that $\OmegaGW$ is constant
for $k>\keq$ and decays as $k^{-2} \propto f^{-2}$ for modes entering
during matter domination.

The same line of reasoning can be applied to the post-inflationary
universe assuming the non-standard history. Just after reheating we
assume the universe to be in a $X$-era (radiation-dominated in our
scenario), then becomes $Y$-dominated, e.g. driven by an oscillating
scalar field, which finally decays into radiation.

\subsubsection{Non-standard post-inflationary eras}

If an observable mode today entered the Hubble radius during the
$Y$-era, Eq.~(\ref{eq:reenter}) can be dealt exactly as the
radiation-era case, provided $\wy$ remains constant. The equivalent of
Eq.~(\ref{eq:ak2}) now reads
\begin{equation}
\ak^2  = \dfrac{3 \Mpl^2 k^2}{\rhoyend \ayend^{3 + 3\wy}} \ak^{3 + 3\wy}\,,
\end{equation}
whose solution can be recast into
\begin{equation}
\label{eq:aky}
\ak = \ayend \left(\dfrac{\kyend}{k} \right)^{2/(1 + 3 \wy)}\,.
\end{equation}
This expression can be further simplified by remarking that 
\begin{equation}
\label{eq:ayend}
\ayend = \azero \rdof_\uyend^{1/4} \left(\dfrac{\rhoradnow}{\rhoyend}
\right)^{1/4} \,,
\end{equation}
since the end of the $Y$-era matches with the beginning of the
standard radiation dominated era. The wavenumber $\kyend$ correspond
to a mode entering the Hubble radius just at the end of the $Y$-era,
i.e. at the beginning of the radiation era:
\begin{equation}
\label{eq:kyend}
\kyend \equiv \dfrac{\ayend}{\sqrt{3} \Mpl} \sqrt{\rhoyend} =
\dfrac{1}{\sqrt{3} \Mpl} (\rhotradnow \rhoyend)^{1/4} \,,
\end{equation}
where the last equality comes from Eq.~(\ref{eq:ayend}). At this
point, Eq.~(\ref{eq:aky}) shows that the quantities $\kyend$ and $\wy$
are observable and completely determined by the measurement of
$\OmegaGW(k)$. As a result, we should try to express all quantities in
terms of them, and in particular the redshift at which the $Y$-era
ended. From Eq.~(\ref{eq:ayend}), one gets
\begin{equation}
\label{eq:zyend}
1+\zyend = \sqrt{3} \, \dfrac{\kyend \Mpl}{\rhotradnow^{1/2}}\,.
\end{equation}
We finally get the transfer function during the $Y$-era
\begin{equation}
\label{eq:ty}
\begin{aligned}
\calT_\uy(k) & \simeq \dfrac{\rhotradnow^{1/2}}{\sqrt{3}\, \kyend
  \Mpl} \left(\dfrac{\kyend}{k}\right)^{2/(1+3\wy)}
\Heaviside{k>\kyend} \\
& + \Heaviside{k<\kyend} .
\end{aligned}
\end{equation}

It remains now to deal with the modes entering the Hubble radius
before, i.e. either during $X$ domination or during reheating. The
calculations are again the same although the scale factor at the end
of the $X$-era reads
\begin{equation}
  \axend = \ayend \dfrac{\axend}{\ayend} = \ayend \Ry \left(
  \dfrac{\rhoyend}{\rhoyini} \right)^{1/4},
\end{equation}
where Eq.~(\ref{eq:Rx}) has been used for $ \axend / \ayend$. One can
further simplify this expression by using Eq.~(\ref{eq:ayend}) to get
\begin{equation}
\label{eq:axend}
\axend = \Ry \left(\dfrac{\rhotradnow}{\rhoxend} \right)^{1/4}\,.
\end{equation}
Defining $\kxend$ the wavenumber of a mode entering the Hubble radius
at the end of the $X$-era, we have
\begin{equation}
\label{eq:kxend}
\kxend \equiv \dfrac{\axend}{\sqrt{3} \Mpl} \sqrt{\rhoxend} =
\dfrac{\Ry}{\sqrt{3} \Mpl} (\rhotradnow \rhoxend)^{1/4} \,,
\end{equation}
such that the corresponding redshift can be expressed in terms of
observable quantities as
\begin{equation}
\label{eq:zxend}
1 + \zxend = \sqrt{3}\, \dfrac{\kxend \Mpl}{\Ry^2 \rhotradnow^{1/2}}\,.
\end{equation}

The transfer function during this era is again given by $|\ak|$ and
reads
\begin{equation}
\label{eq:tx}
\begin{aligned}
\calT_\ux(k) & \simeq \dfrac{\Ry^2
  \rhotradnow^{1/2}}{\sqrt{3}\, \kxend \Mpl}
\left(\dfrac{\kxend}{k}\right)^{2/(1+3\wx)} \Heaviside{k>\kxend} \\ &
+ \Heaviside{k<\kxend}.
\end{aligned}
\end{equation}

Finally, in order to deal with modes entering the Hubble radius during
reheating (inflaton oscillation dominated era), we similarly express
$\areh$ in terms of $\axend$ to get
\begin{equation}
\areh = \Rx \Ry \left(\dfrac{\rhotradnow}{\rhoreh} \right)^{1/4}\,.
\end{equation}
Again, the wavenumber crossing the end of reheating is given by
\begin{equation}
\label{eq:kreh}
\kreh \equiv \dfrac{\areh}{\sqrt{3} \Mpl} \sqrt{\rhoreh} =
\dfrac{\Rx \Ry}{\sqrt{3}\Mpl} (\rhotradnow \rhoreh)^{1/4}\,,
\end{equation}
and the redshift at which reheating ended reads
\begin{equation}
\label{eq:zreh}
1 + \zreh = \sqrt{3}\, \dfrac{\kreh \Mpl}{\Rx^2 \Ry^2 \rhotradnow^{1/2}}\,.
\end{equation}
Therefore, the transfer function during reheating is
\begin{equation}
\label{eq:treh}
\begin{aligned}
\calT_\ureh(k) & \simeq \dfrac{\Rx^2 \Ry^2
  \rhotradnow^{1/2}}{\sqrt{3}\, \kreh \Mpl}
\left(\dfrac{\kreh}{k}\right)^{2/(1+3\wreh)} \Heaviside{k>\kreh} \\ &
+ \Heaviside{k<\kreh},
\end{aligned}
\end{equation}
which shows the influence of $\Rx$ and $\Ry$, making them observable
in the gravitational wave spectrum. However, contrary to CMB, they are
no longer degenerated with $\Rrad$. Conversely, using both CMB and
$\OmegaGW$, one expects to be able to disambiguate the effects of
$\Rrad$, $\Rx$ and $\Ry$. The same reasoning could be extended to
another $Z$-era inserted somewhere.

To summarize, one finally gets
\begin{equation}
  \OmegaGW(k) \simeq \dfrac{k^2}{12 H_0^2} \calT^2_{\ueq}(k) \calT^2_\uy(k)
  \calT^2_\ux(k) \calT^2_\ureh(k) \calP_h(k)\,,
\end{equation}
where the three transfer functions are given by
Eqs.~(\ref{eq:tradmat}), (\ref{eq:ty}), (\ref{eq:tx}) and
(\ref{eq:treh}) and their respective pivot wavenumbers by
Eqs.~(\ref{eq:keq}), (\ref{eq:kyend}), (\ref{eq:kxend}) and
(\ref{eq:kreh}). All of those four wavenumbers therefore correspond to
frequencies of the gravitational wave $\feq$, $\fyend$, $\fxend$ and
$\freh$.

\section{Forecasts on thermal history from future CMB and GW
  experiments}
\label{sec:const}

In this section, we discuss how and to what extent we can probe the
thermal history of the Universe with future CMB and direct detection
of GW experiments.

\subsection{An illustrative scenario}

For illustrative purposes, we consider in the following the specific
scenario mentioned earlier. Inflation is driven by a massive field
$\phi$ having the potential $V(\phi) = (1/2) m^2 \phi^2$ with $m$
being the inflaton mass.  Furthermore there also exists another scalar
field, denoted as $\sigma$ that comes to dominate the universe at
later times. After inflation, $\phi$ oscillates around the minimum of
its potential and the reheating (inflaton oscillation dominated) era
is matter-like~\cite{Turner:1983he} till the Universe thermalizes and
becomes radiation-dominated (the $X$-era). The reheating temperature
$\Treh$, associated with the energy density $\rhoreh$, refers to the
time at which the reheating era ends and the radiation $X$-era
starts. Then the field $\sigma$ begins to oscillate at some epoch and
starts the $\sigma$ oscillation dominated era, referred to as the
$Y$-era in the previous section. We moreover assume that $\sigma$
decays after it dominates the Universe to start the usual radiation
dominated period. Thus, the thermal history of the Universe proceeds
as follows: Inflation $\rightarrow$ reheating era (oscillating $\phi$
dominated era) $\rightarrow$ RD ($X$-era) $\rightarrow$ Oscillating
$\sigma$ dominated era ($Y$-era) $\rightarrow$ RD era. The final RD
era continues until the radiation-matter equality, just before the
recombination epoch. As discussed in the previous section, depending
on the duration of each era, the predictions for CMB and GW spectra
are different, from which we can probe the thermal history.

Under these hypothesis, we can express all the quantities of the
previous section in terms of temperatures and frequencies.  With
$\Rx=1$, assuming that $\gstarE=\gstarS=\gs$ at the time of ``yend,"
``xend," and ``reh" and using Eq.~(\ref{eq:Fy}),
Eqs.~(\ref{eq:kyend}), (\ref{eq:kxend}) and (\ref{eq:kreh}), one
obtains
\begin{equation}
\begin{aligned}
\fyend 
& \simeq 0.2~{\rm Hz} \left( \dfrac{\gsyend}{100} \right)  \left( \dfrac{T_\sigma}{10^7~\GeV} \right), \\
\fxend 
& \simeq 0.2~{\rm Hz} \left( \dfrac{\gsyend}{100} \right)  \left( \dfrac{T_\sigma}{10^7~\GeV} \right) \Fy^{2/3}, \\
\freh 
& \simeq 0.2~{\rm Hz} \left( \dfrac{\gsyend}{100} \right)  \left( \dfrac{\Treh}{10^7~\GeV} \right) \Fy^{-1/3},
\end{aligned}
\label{eq:freqs}
\end{equation}
where $T_\sigma$ is the temperature at which $\sigma$ decays into radiation.

For CMB constraints, one can also simplifies the quantity $\Rrad \Ry$
in terms of temperatures as
\begin{equation}
\begin{aligned}
\ln (\Rrad \Ry)& = \dfrac{1}{3} \ln \dfrac{\Treh}{\Mpl} - \dfrac{1}{3}
\ln \Fy \\ & - \dfrac{1}{12} \ln \dfrac{\rhoend}{\Mpl^4} + \frac{1}{12} \ln
\left( \dfrac{\pi^2}{30} \dfrac{\gsreh \gsyend}{\gsxend} \right),
\label{eq:lnRradRy}
\end{aligned}
\end{equation}
where $\wrehb=0$ has been used. Since $\Treh$ is assumed to be bigger
than $\Tyend$, we have the expected hierarchy $\freh > \fxend >
\fyend$.  For the effective degrees of freedom, we assume those of the
standard model of particle physics, i.e.,
$\gsreh=\gsxend=\gsyend=106.75$ in the following analysis. Notice that
this is certainly not verified but the influence of $\gs$ remains very
small in the final forecasts, the sensible quantities being
logarithmic, see Eq.~(\ref{eq:lnRradRy}).

\begin{figure}
\begin{center}
\includegraphics[width=\linewidth]{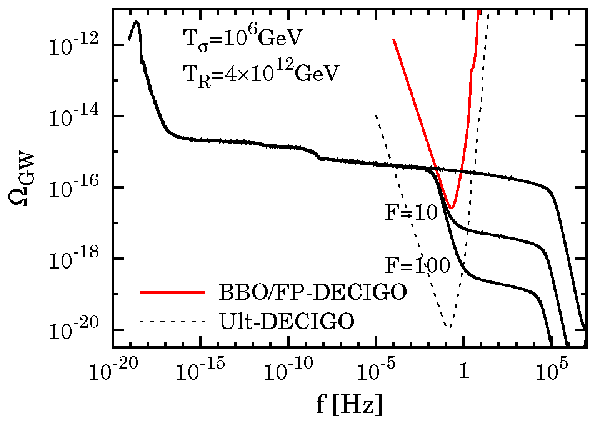}
\includegraphics[width=\linewidth]{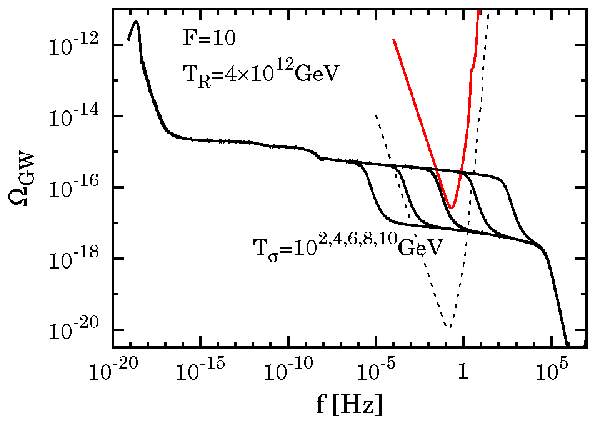}
\caption{GW spectrum for several values of $\Fy$ ranging from $\Fy=0$
  to $\Fy=100$ at fixed $\Tyend$ (top). The bottom panel shows the
  spectra at fixed $\Fy=10$ and for various values of $\Tyend$. The
  reheating temperature $\Treh$ from the inflaton is fixed at $\Treh =
  10^{12}\,\GeV$ in these figures.
\label{fig:Omega_GW}}
\end{center}
\end{figure}

Some example plots for GW spectrum are shown in
Fig.~\ref{fig:Omega_GW}.  The GW spectrum is obtained by numerically
integrating Eq.~(\ref{eq:tensevol}), using the WKB approximation
($h\propto e^{\pm ik\eta}/a$) for the oscillating phase.  To see how
the spectrum shape depends on the reheating temperatures and $F$, we
show several cases for these parameters. For reference, the
sensitivity frequency bands for BBO/FP-DECIGO and Ultimate DECIGO are
also depicted (for the specifications of these experiments, see
Table~\ref{GWparm}).  In some cases, the change of the spectrum can be
traced with future experiments, in particular, for Ultimate DECIGO.
In such a case, the parameters such as the reheating temperatures and
$F$ can be well determined, which is going to be studied by using
Fisher matrix analysis in the following.

\subsection{Fisher matrix analysis}

To forecast constraints from future experiments, we use a Fisher
matrix analysis for both CMB~\cite{Zaldarriaga:1997ch} and GW direct
detection~\cite{Seto:2005qy}.  Under the assumption of a Gaussian
likelihood, the Fisher matrix is given by the second derivative of the
log-likelihood with respect to the parameters $p_i$ at the likelihood
maximum,
\begin{equation}
{\cal F}_{ij}=-\left\langle\frac{\partial^2 \ln{\cal L}}
{\partial p_i \partial p_j}\right\rangle,
\end{equation}  
and its inverse gives marginalized $1\sigma$ error of the parameter
of interest,
\begin{equation}
     \sigma(p_{i}) =\sqrt{({\cal F}^{-1})_{ii}} \,. 
\end{equation}

\subsubsection{CMB experiment}

In the case of CMB experiment, we assume that the Fisher matrix for
both CMBpol and PLANCK is given by
\begin{equation}
\calF_{ij}=\sum_{\ell=2}^{\ell_{\max}} \sum_{XX',YY'}
\frac{\partial C_{\ell}^{XX'}}{\partial p_i} (\Cov_{\ell}^{-1})_{XX'YY'}\frac{\partial C_{\ell}^{YY'}}{\partial
  p_j},
\end{equation}
where $X$ and $X'$ are summed over the CMB temperature, $E$-mode
polarization, and $B$-mode polarization ($X=T,E,B$). The covariance
matrix is given by
\begin{equation}
\begin{aligned}
(\Cov^{-1})_{TTTT} &=\frac{2}{(2l+1)f_{\rm
      sky}}(C_{\ell}^{TT}+w_T^{-1}B_\ell^{-2})^2,
  \\ (\Cov^{-1})_{EEEE} &=\frac{2}{(2l+1)f_{\rm
      sky}}(C_{\ell}^{EE}+w_P^{-1}B_\ell^{-2})^2,
  \\ (\Cov^{-1})_{BBBB} &=\frac{2}{(2l+1)f_{\rm
      sky}}(C_{\ell}^{BB}+w_P^{-1}B_\ell^{-2})^2,
  \\ (\Cov^{-1})_{TETE} &=\frac{2}{(2l+1)\fsky}[(C_{\ell}^{TE})^2
    \\ +(C_{\ell}^{TT} &+
    w_T^{-1}B_\ell^{-2})(C_{\ell}^{EE}+w_P^{-1}B_\ell^{-2})],
  \\ (\Cov^{-1})_{TTEE} &=\frac{2}{(2l+1)\fsky}(C_{\ell}^{TE})^2,
  \\ (\Cov^{-1})_{TTTE} &=\frac{2}{(2l+1)f_{\rm
      sky}}C_{\ell}^{TE}(C_{\ell}^{TT}+w_T^{-1}B_\ell^{-2})^2,
  \\ (\Cov^{-1})_{EETE}
  &=\frac{2}{(2l+1)\fsky}C_{\ell}^{TE}(C_{\ell}^{EE}+w_P^{-1}B_\ell^{-2})^2,
\end{aligned}
\end{equation}
where $w_{(T,P)}^{-1}=4\pi\sigma_{(T,P)}^2/\Npix$ is the variance of
the noise temperature per pixel (in $\muK^2$). For simplicity, we have
assumed a Gaussian beam $B_\ell \simeq
\exp[-\ell(\ell+1)\sigma_\ub^2/2]$ with $\sigma_\ub=\theta/\sqrt{8 \ln
  2}$ being the beam width. In Table~\ref{CMBparm}, we list the values
of the observed fraction of the sky $\fsky$, the temperature noise per
pixel $\sigma_T$, the polarization noise per pixel $\sigma_P$ and the
Gaussian beam width $\theta^2=4\pi/\Npix$ for Planck \cite{:2006uk}
and CMBpol \cite{Baumann:2008aq} experiments that are adopted to
derive our forecasts. The different frequency channels are combined
according to $w_{(T,P)} B_\ell^2=\sum_\nu w_{(T,P)}^\nu
(B_\ell^{\nu})^2$, where $\nu$ refers to each channel
component~\cite{1997MNRAS.291L..33B}. Finally, the maximum multipole
value has been set to $\ell_{\max} = 2000$.

\begin{table*}
\begin{center}
\caption{Instrument parameter values for CMB experiments 
\label{CMBparm}}
\begin{tabular*}{0.7\textwidth}{@{\extracolsep{\fill}}lcccccc}
\hline
Experiment&$\fsky$&Center frequency [GHz]&$\theta$ [FWHM arcmin]&$\sigma_T$ [$\mu{\rm K}$]&$\sigma_P$ [$\mu{\rm K}$]\\
\hline
\hline
Planck \cite{:2006uk} &0.65&70&14&12.8&18.2\\
&&100&10&6.8&10.9\\
&&143&7.1&6.0&11.4\\
&&217&5.0&13.1&26.7\\
\hline
CMBpol \cite{Baumann:2008aq} &0.65&100&4.2&0.87&1.18\\
&&150&2.8&1.26&1.76\\
&&220&1.9&1.84&2.60\\
\hline
\end{tabular*}
\end{center}
\end{table*}

Hereafter, we assume a flat $\Lambda$CDM Universe and set the fiducial
cosmological parameters to the WMAP7 mean 
values~\cite{Komatsu:2010fb}.

\subsubsection{GW direct detection}

For the GW direct detection experiments, we will be considering three
future experiments, FP-DECIGO, BBO and Ultimate-DECIGO \cite{
  Kawamura:2011zz,Harry:2006fi,Seto:2001qf}.  FP-DECIGO is planned to
be a Fabry-Perot Michelson interferometer, while BBO experiment will
use time-delay interferometry (TDI).  Although they use different
technology, in this paper, we do not distinguish these experiments,
since they have similar sensitivity.  The Fisher matrix for GW direct
detection is given by \cite{Seto:2005qy}
\begin{equation}
\begin{aligned}
\calF_{ij} & = \left(\frac{3H_0^2}{10\pi^2}\right)^2 2\, \tobs \\
& \times \sum_{(I,J)} \int^{\infty}_{0} \ud f \frac{|\gamma_{IJ}(f)|^2\partial_{p_i}
     \OmegaGW(f) \partial_{p_j} \OmegaGW(f)}{f^6 S_I(f) S_J(f)},
\end{aligned}
\label{Fisher}
\end{equation}
where $\OmegaGW$ is given in Eq.~(\ref{eq:omegaGW}) and $\tobs$ is the
observation time. The subscripts $I$ and $J$ refer to independent
signals obtained at each detector, or observables generated by
combining the detector signals. For a BBO-like experiment, the the
summation runs over the TDI channel output index ($I=A,E,T$)
\cite{Seto:2005qy}.  The overlap reduction $\gamma_{IJ}(f)$
for TDI variables is calculated following the procedure of
Ref.~\cite{Corbin:2005ny}. 
For the cross term ($I\neq J$), we have $\gamma_{IJ}(f)=0$. 
The noise spectrum $S_{I,J}(f)$ is given by
\begin{equation}
\begin{aligned}
S_A(f) & = S_E(f) = 8 \sin^2(\hat{f}/2) \left\{ (2+\cos\hat{f}) \Sshot
\right. \\ & + \left. 2 \left[ 3 + 2\cos\hat{f} + \cos(2\hat{f})
  \right] \Saccel \right\}, \\ S_T(f) &= 2 \left( 1 +
2\cos\hat{f}\right)^2 \left[\Sshot + 4\sin^2(\hat{f}/2) \Saccel
  \right],
\end{aligned}
\end{equation}
where $\hat{f}=2\pi Lf$.  The values of the arm length $L$, the shot
noise $\Sshot$ and the radiation pressure noise $\Saccel$ have been
reported in Table~\ref{GWparm} for BBO/FP-DECIGO and Ultimate-DECIGO,
respectively. One may be concerned about the noise contamination from
white dwarf binaries at frequencies below $\sim 0.1\,\Hz$, and
introduce a low-frequency cutoff to the integral. However, it may be
possible to remove it by identifying all binaries and subtracting
their contributions from data streams~\cite{Cutler:2005qq,
  Harms:2008xv, Yagi:2011wg}. In this paper, we assume that this is
the case and do not introduce such a low-frequency cutoff.

Although we focus on the inflationary GW background in this paper, it
should also be noted that there may be other possible GW signals in
the sensitivity range of BBO and DECICO. Such examples include for
instance first order phase transition~\cite{Kosowsky:1992rz,
  Kosowsky:1991ua, Kamionkowski:1993fg, Caprini:2007xq,
  Caprini:2009yp}, preheating~\cite{Khlebnikov:1997di, Easther:2006gt,
  GarciaBellido:2007dg, GarciaBellido:2007af, Dufaux:2007pt}, particle
production~\cite{Barnaby:2012xt} and cosmic
strings~\cite{Vilenkin:1981bx,Hogan:1984is,Vachaspati:1984gt,Damour:2001bk,
  Olmez:2010bi}. These GWs may be significant in some cases but they
will not be considered in the following.

\begin{table*}
\begin{center}
\caption{Instrument parameters for GW experiments 
\label{GWparm}}
\begin{tabular*}{0.8\textwidth}{@{\extracolsep{\fill}}cccc}
  \hline 
  Experiments
  &$\Sshot$ $[(L/{\rm km})^{-2}{\rm Hz}^{-1}]$
  &$\Saccel$ $[(2\pi f/{\rm Hz})^{-4}(2 L/{\rm km})^{-2}{\rm Hz}^{-1}]$
  &$L$ [km]\\
  \hline
  \hline
  BBO/FP-DECIGO \cite{Corbin:2005ny}
  & $2\times 10^{-40}$
  & $9\times 10^{-40}$
  & $5\times 10^{4}$\\
  \hline
  Ultimate DECIGO \cite{Kudoh:2005as}
  & $9\times 10^{-44}$
  & $9\times 10^{-44}$
  & $5\times 10^{4}$\\
  \hline 
\end{tabular*}
\end{center}
\end{table*}

\subsection{Forecasts on early Universe history}

Now we investigate the constraint on the thermal history of the
Universe using the Fisher matrix method discussed in the previous
section. For this purpose, we focus on the reheating temperature
$\Treh$ associated with the inflaton, the temperature associated with
the second field $\Tyend$, and the amount of late-time entropy
production $\Fy$ (or equivalently $\Ry$).

\subsubsection{Unobservable spectral features}

In this section, we have performed a Fisher analysis to obtain the
expected constraints from Planck/CMBpol for CMB and BBO/DECIGO for GW
direct detection. As a fiducial model for the analysis, we consider
two cases with $\gsreh^{1/4} \Treh \sim 10^{16}\,\GeV$ (corresponding
to $N \equiv - \Nk \sim 57$) and $\gsreh^{1/4} \Treh \sim
10^{9}\,\GeV$ (corresponding to $N\sim52$), assuming that there is no
late-time entropy production, i.e., $\Fy=1$.

\begin{figure}[t]
\begin{center}
\includegraphics[width=\linewidth]{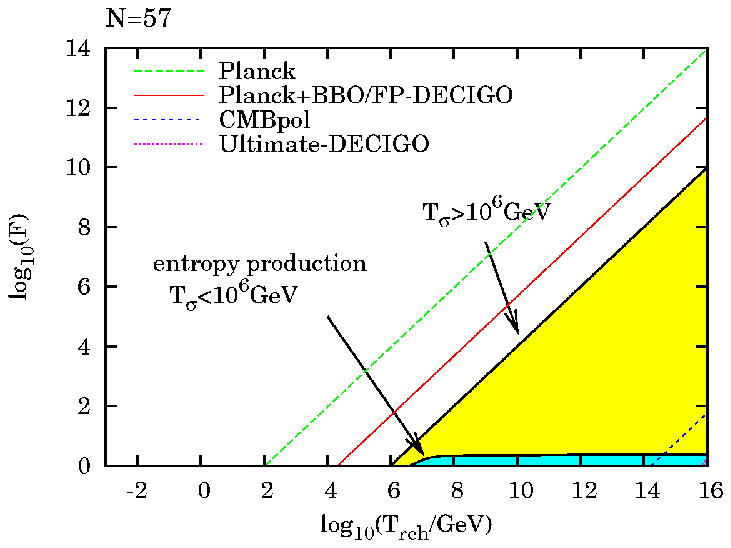}
\includegraphics[width=\linewidth]{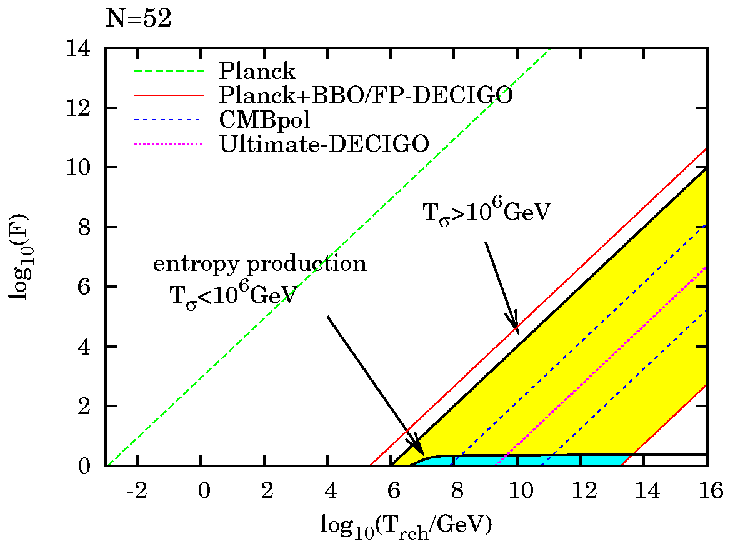}
\caption{Future constraints on the $(\Treh,\Fy)$ plane from CMB
  (Planck and CMBpol) and/or direct detection of GW (BBO/FP-DECIGO and
  Ultimate DECIGO). The light shaded region traces the 2$\sigma$
  confidence interval of the two-dimensional marginalized posterior
  under the Fisher matrix analysis. For the darker blue shaded region
  traces the parameter space in which the signal-to-noise ratio
  $S/N>5$ (see text).}
\label{fig:TR_F}
\end{center}
\end{figure}

\begin{figure}[t]
  \begin{center}
    \includegraphics[width=\linewidth]{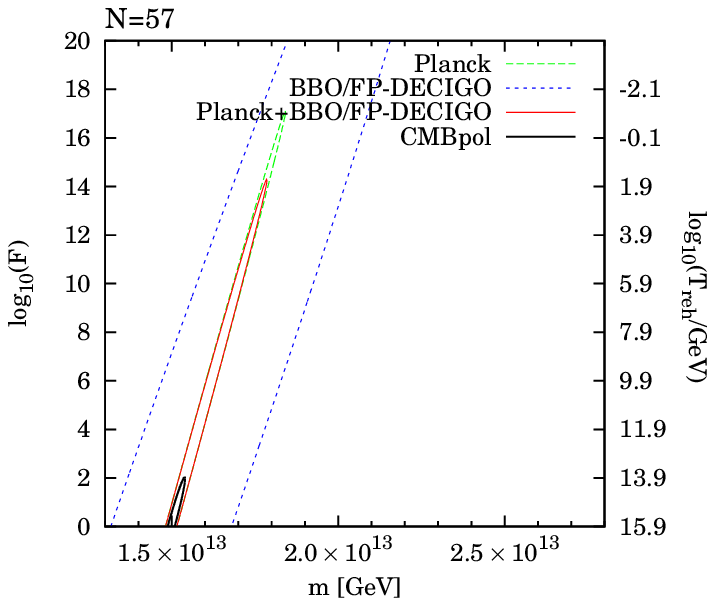}
    \includegraphics[width=\linewidth]{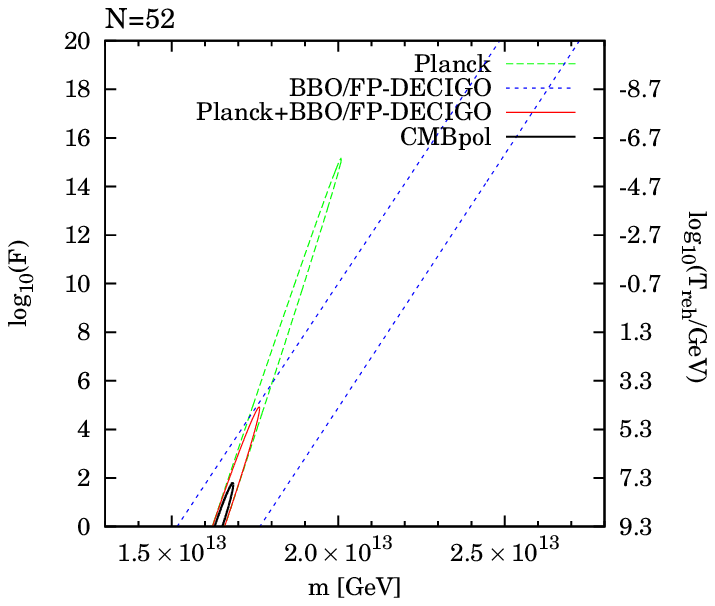}
    \caption{Future constraint in the plane $(m,\Fy)$ from CMB (Planck
      and CMBpol) and GW (BBO/FP-DECIGO), for the two fiducial models
      having $\Treh \simeq 10^{9}\,\GeV$ ($N=52$) and $\Treh \simeq
      10^{16} \,\GeV$ ($N=57$).}
    \label{fig:m_F}
  \end{center}
\end{figure}

For the above fiducial models, there is no spectral signatures in the
sensitivity range of the GWs direct detection experiments and the
Fisher analysis remain insensitive to any information coming from the
spectral shapes. As a result, one expects some degeneracies to occur
between the model parameters. In Figs.~\ref{fig:TR_F} and
\ref{fig:m_F}, we have represented the $2 \sigma$ allowed regions in
the plane $(\Treh,\Fy)$ and $(m,\Fy)$, respectively, for several
combinations of those future data.

There are two cases to consider according to the values of
$\Tyend$. In Fig.~\ref{fig:TR_F}, we have also represented the region
where the signal-to-noise ratio is $S/N > 5$ for BBO/FP-DECIGO as the
light blue shaded region. As studied in Fig.~\ref{fig:Omega_GW}, a
late time entropy production would induce a suppression of the GW
spectrum amplitude at frequencies higher than $\fyend$, which
corresponds to $\Tyend$ in Eq.~(\ref{eq:freqs}). For BBO/FP-DECIGO,
the suppression occurs in the sensitivity region when $\Tyend <
10^{6}\,\GeV$. In that situation, a direct detection of the primordial
GWs by BBO/FP-DECIGO would immediately yield a strong upper bound for
$\Fy \lesssim 2.4$ (see Fig.~\ref{fig:TR_F}). At the same time, since the
reheating also induces a suppression of the GWs spectrum amplitude,
direct detection would readily excludes $\Treh < 10^{6}\, \GeV$ for
the same reason.

The opposite situation, i.e. the case with $\Tyend > 10^{6}\, \GeV$,
the spectrum is suppressed at higher frequencies and beyond the
sensitivity range of the experiment. Therefore, the amplitude of the
observable GWs spectrum remains mostly independent of the values of
$\Fy$. In this case, direct detection of GWs still allows a large
parameter space represented as the light yellow shading in
Fig.~\ref{fig:TR_F}. Within this region, the upper limit of $\Fy$,
equivalently the lower limit of $\Treh$, is imposed by the
consideration that reheating should end before the late-time entropy
production begins.  In other words, $\fxend<\freh$ gives $\Tyend
\Fy^{2/3}<\Treh \Fy^{-1/3}$ using Eq. (\ref{eq:freqs}), and given
$\Tyend > 10^{6} \, \GeV$, we get $\Treh/\Fy>10^{6}\,\GeV$. Since the
constraints on $\Fy$ are completely degenerated with the one on
$\Treh$, we have also reported the values of $\Treh$ along the right
vertical axis of Fig.~\ref{fig:m_F}.

Regarding CMB constraints, there is also a degeneracy between $\Treh$
and $\Fy$, which is clear from Eq. (\ref{eq:lnRradRy}). However, the
inflaton mass $m$ can be probed through the amplitude of primordial
curvature perturbation while $\nS$ and $r$ are related to $\Fy$ and
$\Treh$. On the other hand, the constraints from GW experiments
basically come from their sensitivity to $r$ and $\nT$.  Since in the
single field inflation model, these two quantities are related by $\nT
= -r/8$ (at leading order in slow-roll parameters), $m$ and $\Fy$ are
constrained in a different way compared to CMB. In addition, CMB and
GW direct detection experiments measure slow-roll parameters
quantities at different scales. As a result, the direction of this
degeneracy differs thereby showing the complementarity of these
observables.

\subsubsection{Detection of spectral features}

For Ultimate-DECIGO, the suppression region of the GW signal occurs at
$\Tyend < 10^3 \, \GeV$. However, there are large possibilities that
the extreme sensitivity of Ultimate-DECIGO enables us to measure some
of the spectral features. In that case, the detection of GWs would
provide a precise determination of the
parameters~\cite{Kuroyanagi:2011fy}. In the following, we investigate
in more details the determination of the thermal history parameters by
Ultimate-DECIGO in combination with CMBpol.

\begin{figure}
\begin{center}
\includegraphics[width=\linewidth]{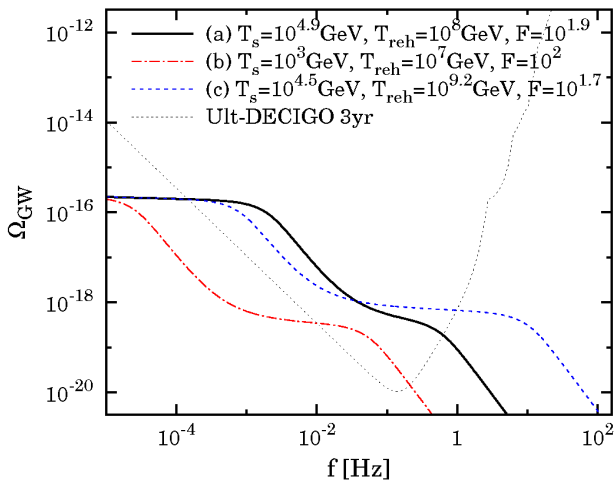}
\caption{Spectra of the three fiducial models (a), (b) and (c)
  together with the sensitivity region of the Ultimate-DECIGO GW
  experiment (see text). The corresponding forecasts are represented
  in Fig.~\ref{fig:constraintsUlt}.}
\label{fig:spectrumUlt}
\end{center}
\end{figure}

\begin{figure}
\begin{center}
\includegraphics[width=0.82\linewidth]{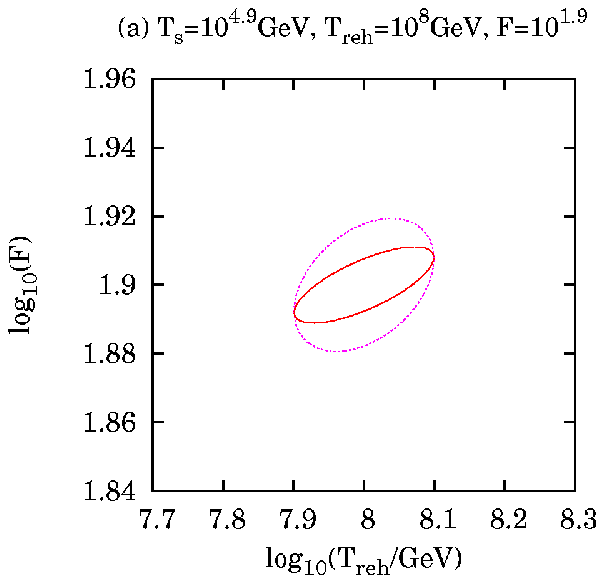}
\includegraphics[width=0.82\linewidth]{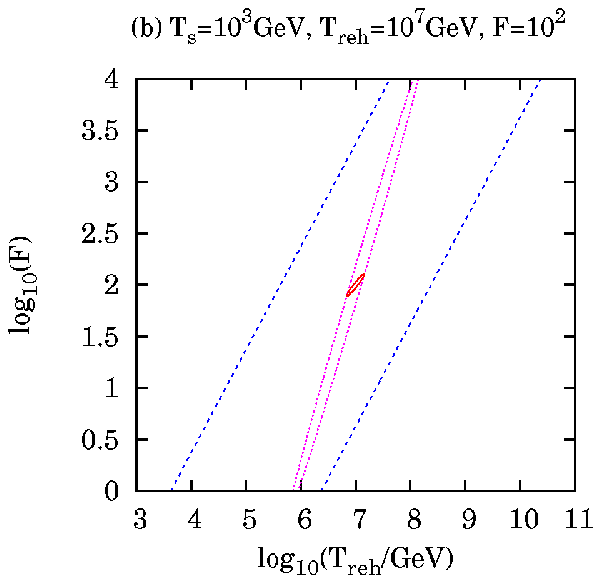}
\includegraphics[width=0.82\linewidth]{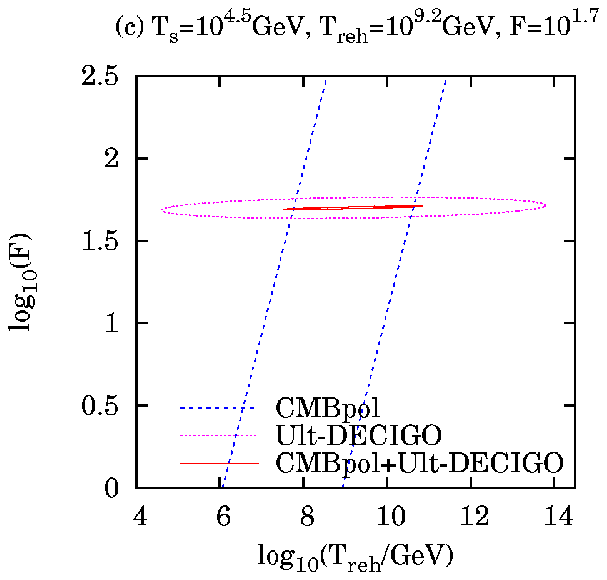}
\caption{Predicted constraints in the $(\Treh,\Fy)$ plane for the
  three different fiducial models (a), (b) and (c). The dotted and
  dashed lines show the marginalized $2\sigma$ confidence contours
  expected from Ultimate-DECIGO and CMBpol, respectively. The solid
  line is the combined constraints. See also
  Fig.~\ref{fig:constraintsUltmF}.}
\label{fig:constraintsUlt}
\end{center}
\end{figure}

In Fig.~\ref{fig:constraintsUlt}, we show the constraints on the
parameters $\Treh$ and $\Fy$ expected from Ultimate-DECIGO together with
ones expected from CMBpol. Three different fiducial models have been
considered:
\begin{itemize}
\item (a) Both transition frequencies are inside the sensitivity
  frequency band of the experiment. We use the fiducial values $\Tyend
  = 10^{4.9} \, \GeV$, $\Treh=10^8\,\GeV$ and $\Fy=10^{1.9}$.
\item (b) The frequency $\fxend$ is outside the sensitivity
  range. As such an example, we consider the case with $\Tyend=10^3\,\GeV$,
  $\Treh=10^7\,\GeV$ and $\Fy=10^2$.
\item (c) The frequency $\freh$ is outside the sensitivity range,
  which occurs for the fiducial values $\Tyend=10^{4.5} \, \GeV$,
  $\Treh=10^{9.2} \, \GeV$ and $\Fy=10^{1.7}$.
\end{itemize}
The spectra of these three models have been represented in
Fig.~\ref{fig:spectrumUlt}.  Here, we take $\Treh$, $\Fy$, $\Tyend$
and $m$ as free parameters and the predicted constraints are obtained
by marginalizing over the remaining parameters. For CMBpol, the other
cosmological parameters are also marginalized.

\begin{figure}
\begin{center}
\includegraphics[width=0.82\linewidth]{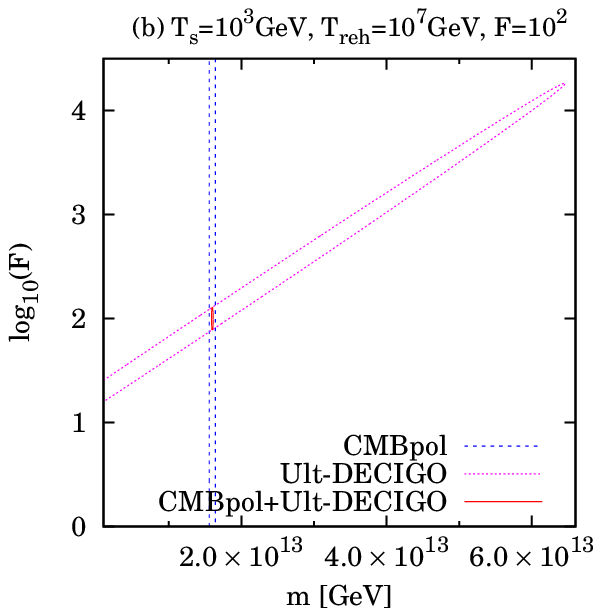}
\caption{Predicted constraints for the fiducial model (b) in the
  $(m,\Fy)$ plane for Ultimate-DECIGO, CMBpol and both (same
  convention as in Fig.~\ref{fig:constraintsUlt}).}
\label{fig:constraintsUltmF}
\end{center}
\end{figure}

\begin{figure}
\begin{center}
\includegraphics[width=0.82\linewidth]{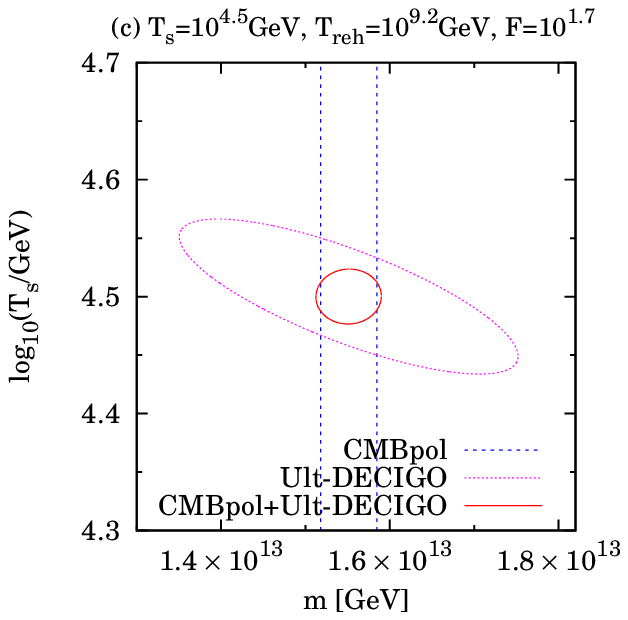}
\caption{Predicted constraints for the fiducial model~(c) in the
  $(m,\Tyend)$ plane for Ultimate-DECIGO, CMBpol and both (same
  convention as in Fig.~\ref{fig:constraintsUlt}).}
\label{fig:constraintsUltmTs}
\end{center}
\end{figure}

In the case (a), all features of the reheating and late-time entropy
production are within the GW sensitivity range. Since, in this case,
Ultimate-DECIGO alone can well determine all the parameters, we do not
show the constraint from CMBpol alone (in fact, the $1\sigma$
contour from CMBpol ends up being outside the range of the figure).
However, it should be noted here that adding CMB data improves the
constraint, as seen from the figure.

For the case (b), the damping of the GW spectrum within the
sensitivity zone is due to the inflationary reheating and $\Treh$ can
be inferred. On the other hand, we find that GW measurements alone
cannot determine the value of $\Fy$, since the feature of late-time
entropy production is now outside the region of detectability.
However, if we combine CMB data, the constraint can be improved as
seen from second panel of Fig.~\ref{fig:constraintsUlt}. To illustrate
this more clearly, we show in Fig. \ref{fig:constraintsUltmF} the
marginalized constraints for case (b) in the $(m,\Fy)$
plane. Constraints from direct detection have a strong parameter
degeneracy between $m$ and $\Fy$, since they both cause suppression of
the amplitude of GWs at the maximum sensitivity region of
Ultimate-DECIGO.  However, CMB data such as CMBpol are strongly
sensitive to the value of $m$ and greatly helps to break the
degeneracy. Therefore, although CMBpol data alone only provides
constraints on $\Fy$ and $\Treh$ with large uncertainties (see middle
panel of Fig.~\ref{fig:constraintsUlt}), it can significantly tighten
the constraints from Ultimate-DECIGO through improving the
determination on $m$.

In the case (c), the reheating frequency is outside the sensitivity
region such that the value of $\Treh$ can no longer be well determined
by GW measurements. Still $\Tyend$ and $\Fy$ can be well probed
instead. As a result, the constraint from Ultimate DECIGO lies
parallel to the $\Treh$ axis (see
Fig.~\ref{fig:constraintsUlt}). However, as repeatedly emphasized
above, CMB probes the parameters differently from GW, thus in
combination with CMBpol, $\Treh$ can again be inferred and this ends
up with breaking its degeneracy with $\Fy$. Furthermore, in
Fig.~\ref{fig:constraintsUltmTs}, we have also depicted the expected
constraints in the $(m,\Tyend)$ plane, which also clearly illustrates
the complementarity of observations of CMB and GW.

\section{Summary}

We have discussed how one can probe the thermal history of the
Universe from the era just after the end of inflation until the BBN
epoch. In any given inflationary models, the spectral index and the
tensor-to-scalar ratio are related to the $e$-fold number at which a
reference scale exited the Hubble radius during inflation, thus in
turn, they give a constraint on the total expansion of the scale
factor since the above-mentioned epoch. By assuming chaotic inflation,
we have presented how the CMB is constraining the thermal history when
there is an epoch of late-time entropy production, as for instance, a
scenario where an oscillating scalar field dominates the Universe at
some epoch, and then it decays.

In the future, direct detection GW experiments are expected to provide
new cosmological probes, in addition to more precise CMB experiments
such as Planck and CMBpol. Although CMB data can be used to determine
the total amount of the cosmic expansion, they are only sensitive to
the integrated thermal history and an epoch of late-time entropy
production remains fully degenerated with the standard inflationary
reheating era. This is not the case for GW experiments as they are
precisely sensitive to the transition between these epochs. In
particular, for the scenario with late-time entropy production, the GW
spectrum bends twice at some transition frequencies (see
Figs.~\ref{fig:Omega_GW} and \ref{fig:spectrumUlt}). We have shown
that if one, or more, frequencies do not lie in the GW experiment
sensitivity region, CMB experiments can still greatly help to break
the degeneracy. As shown in this paper, future experiments of GW and
CMB are complementary in performing a tomography of the thermal
history between the end of inflation and BBN. As such, they are
expected to play a major role in our understanding of the whole
history of the Universe.

\begin{acknowledgments}

  T.T. would like to thank CP3 at Louvain University for the
  hospitality during the visit where a part of the work has been
  done. This work is partially supported by the Grant-in-Aid for
  Scientific research from the Ministry of Education, Science, Sports,
  and Culture, Japan, Nos.~23340058 (S.K.)  and 24740149 (S.K.) and
  23740195 (T.T).  C.R. is partially supported by the ESA Belgian
  Federal PRODEX Grant No.~4000103071 and the Wallonia-Brussels
  Federation grant ARC No.~11/15-040.
\end{acknowledgments}

\bibliography{bibgw,bibinf}

\begin{thebibliography}{55}
\expandafter\ifx\csname natexlab\endcsname\relax\def\natexlab#1{#1}\fi
\expandafter\ifx\csname bibnamefont\endcsname\relax
  \def\bibnamefont#1{#1}\fi
\expandafter\ifx\csname bibfnamefont\endcsname\relax
  \def\bibfnamefont#1{#1}\fi
\expandafter\ifx\csname citenamefont\endcsname\relax
  \def\citenamefont#1{#1}\fi
\expandafter\ifx\csname url\endcsname\relax
  \def\url#1{\texttt{#1}}\fi
\expandafter\ifx\csname urlprefix\endcsname\relax\def\urlprefix{URL }\fi
\providecommand{\bibinfo}[2]{#2}
\providecommand{\eprint}[2][]{\url{#2}}

\bibitem[{\citenamefont{Martin and Ringeval}(2006)}]{Martin:2006rs}
\bibinfo{author}{\bibfnamefont{J.}~\bibnamefont{Martin}} \bibnamefont{and}
  \bibinfo{author}{\bibfnamefont{C.}~\bibnamefont{Ringeval}},
  \bibinfo{journal}{JCAP} \textbf{\bibinfo{volume}{0608}}, \bibinfo{pages}{009}
  (\bibinfo{year}{2006}), \eprint{astro-ph/0605367}.

\bibitem[{\citenamefont{Ringeval}(2008)}]{Ringeval:2007am}
\bibinfo{author}{\bibfnamefont{C.}~\bibnamefont{Ringeval}},
  \bibinfo{journal}{Lect. Notes Phys.} \textbf{\bibinfo{volume}{738}},
  \bibinfo{pages}{243} (\bibinfo{year}{2008}), \eprint{astro-ph/0703486}.

\bibitem[{\citenamefont{Martin and Ringeval}(2010)}]{Martin:2010kz}
\bibinfo{author}{\bibfnamefont{J.}~\bibnamefont{Martin}} \bibnamefont{and}
  \bibinfo{author}{\bibfnamefont{C.}~\bibnamefont{Ringeval}},
  \bibinfo{journal}{Phys. Rev.} \textbf{\bibinfo{volume}{D82}},
  \bibinfo{pages}{023511} (\bibinfo{year}{2010}), \eprint{1004.5525}.

\bibitem[{\citenamefont{Seto and Yokoyama}(2003)}]{Seto:2003kc}
\bibinfo{author}{\bibfnamefont{N.}~\bibnamefont{Seto}} \bibnamefont{and}
  \bibinfo{author}{\bibfnamefont{J.}~\bibnamefont{Yokoyama}},
  \bibinfo{journal}{J.Phys.Soc.Jap.} \textbf{\bibinfo{volume}{72}},
  \bibinfo{pages}{3082} (\bibinfo{year}{2003}), \eprint{gr-qc/0305096}.

\bibitem[{\citenamefont{Nakayama
  et~al.}(2008{\natexlab{a}})\citenamefont{Nakayama, Saito, Suwa, and
  Yokoyama}}]{Nakayama:2008ip}
\bibinfo{author}{\bibfnamefont{K.}~\bibnamefont{Nakayama}},
  \bibinfo{author}{\bibfnamefont{S.}~\bibnamefont{Saito}},
  \bibinfo{author}{\bibfnamefont{Y.}~\bibnamefont{Suwa}}, \bibnamefont{and}
  \bibinfo{author}{\bibfnamefont{J.}~\bibnamefont{Yokoyama}},
  \bibinfo{journal}{Phys.Rev.} \textbf{\bibinfo{volume}{D77}},
  \bibinfo{pages}{124001} (\bibinfo{year}{2008}{\natexlab{a}}),
  \eprint{0802.2452}.

\bibitem[{\citenamefont{Nakayama
  et~al.}(2008{\natexlab{b}})\citenamefont{Nakayama, Saito, Suwa, and
  Yokoyama}}]{Nakayama:2008wy}
\bibinfo{author}{\bibfnamefont{K.}~\bibnamefont{Nakayama}},
  \bibinfo{author}{\bibfnamefont{S.}~\bibnamefont{Saito}},
  \bibinfo{author}{\bibfnamefont{Y.}~\bibnamefont{Suwa}}, \bibnamefont{and}
  \bibinfo{author}{\bibfnamefont{J.}~\bibnamefont{Yokoyama}},
  \bibinfo{journal}{JCAP} \textbf{\bibinfo{volume}{0806}}, \bibinfo{pages}{020}
  (\bibinfo{year}{2008}{\natexlab{b}}), \eprint{0804.1827}.

\bibitem[{\citenamefont{Kuroyanagi et~al.}(2010)\citenamefont{Kuroyanagi,
  Gordon, Silk, and Sugiyama}}]{Kuroyanagi:2009br}
\bibinfo{author}{\bibfnamefont{S.}~\bibnamefont{Kuroyanagi}},
  \bibinfo{author}{\bibfnamefont{C.}~\bibnamefont{Gordon}},
  \bibinfo{author}{\bibfnamefont{J.}~\bibnamefont{Silk}}, \bibnamefont{and}
  \bibinfo{author}{\bibfnamefont{N.}~\bibnamefont{Sugiyama}},
  \bibinfo{journal}{Phys. Rev.} \textbf{\bibinfo{volume}{D81}},
  \bibinfo{pages}{083524} (\bibinfo{year}{2010}), \eprint{0912.3683}.

\bibitem[{\citenamefont{Durrer and Hasenkamp}(2011)}]{Durrer:2011bi}
\bibinfo{author}{\bibfnamefont{R.}~\bibnamefont{Durrer}} \bibnamefont{and}
  \bibinfo{author}{\bibfnamefont{J.}~\bibnamefont{Hasenkamp}},
  \bibinfo{journal}{Phys.Rev.} \textbf{\bibinfo{volume}{D84}},
  \bibinfo{pages}{064027} (\bibinfo{year}{2011}), \eprint{1105.5283}.

\bibitem[{\citenamefont{Komatsu et~al.}(2011)}]{Komatsu:2010fb}
\bibinfo{author}{\bibfnamefont{E.}~\bibnamefont{Komatsu}} \bibnamefont{et~al.}
  (\bibinfo{collaboration}{WMAP}), \bibinfo{journal}{Astrophys. J. Suppl.}
  \textbf{\bibinfo{volume}{192}}, \bibinfo{pages}{18} (\bibinfo{year}{2011}),
  \eprint{1001.4538}.

\bibitem[{\citenamefont{Harry et~al.}(2006)\citenamefont{Harry, Fritschel,
  Shaddock, Folkner, and Phinney}}]{Harry:2006fi}
\bibinfo{author}{\bibfnamefont{G.}~\bibnamefont{Harry}},
  \bibinfo{author}{\bibfnamefont{P.}~\bibnamefont{Fritschel}},
  \bibinfo{author}{\bibfnamefont{D.}~\bibnamefont{Shaddock}},
  \bibinfo{author}{\bibfnamefont{W.}~\bibnamefont{Folkner}}, \bibnamefont{and}
  \bibinfo{author}{\bibfnamefont{E.}~\bibnamefont{Phinney}},
  \bibinfo{journal}{Class.Quant.Grav.} \textbf{\bibinfo{volume}{23}},
  \bibinfo{pages}{4887} (\bibinfo{year}{2006}).

\bibitem[{\citenamefont{Kawamura et~al.}(2011)\citenamefont{Kawamura, Ando,
  Seto, Sato, Nakamura et~al.}}]{Kawamura:2011zz}
\bibinfo{author}{\bibfnamefont{S.}~\bibnamefont{Kawamura}},
  \bibinfo{author}{\bibfnamefont{M.}~\bibnamefont{Ando}},
  \bibinfo{author}{\bibfnamefont{N.}~\bibnamefont{Seto}},
  \bibinfo{author}{\bibfnamefont{S.}~\bibnamefont{Sato}},
  \bibinfo{author}{\bibfnamefont{T.}~\bibnamefont{Nakamura}},
  \bibnamefont{et~al.}, \bibinfo{journal}{Class.Quant.Grav.}
  \textbf{\bibinfo{volume}{28}}, \bibinfo{pages}{094011}
  (\bibinfo{year}{2011}).

\bibitem[{\citenamefont{Seto et~al.}(2001)\citenamefont{Seto, Kawamura, and
  Nakamura}}]{Seto:2001qf}
\bibinfo{author}{\bibfnamefont{N.}~\bibnamefont{Seto}},
  \bibinfo{author}{\bibfnamefont{S.}~\bibnamefont{Kawamura}}, \bibnamefont{and}
  \bibinfo{author}{\bibfnamefont{T.}~\bibnamefont{Nakamura}},
  \bibinfo{journal}{Phys.Rev.Lett.} \textbf{\bibinfo{volume}{87}},
  \bibinfo{pages}{221103} (\bibinfo{year}{2001}), \eprint{astro-ph/0108011}.

\bibitem[{\citenamefont{Schwarz et~al.}(2001)\citenamefont{Schwarz,
  Terrero-Escalante, and Garcia}}]{Schwarz:2001vv}
\bibinfo{author}{\bibfnamefont{D.~J.} \bibnamefont{Schwarz}},
  \bibinfo{author}{\bibfnamefont{C.~A.} \bibnamefont{Terrero-Escalante}},
  \bibnamefont{and} \bibinfo{author}{\bibfnamefont{A.~A.}
  \bibnamefont{Garcia}}, \bibinfo{journal}{Phys. Lett.}
  \textbf{\bibinfo{volume}{B517}}, \bibinfo{pages}{243} (\bibinfo{year}{2001}),
  \eprint{astro-ph/0106020}.

\bibitem[{\citenamefont{Kuroyanagi and Takahashi}(2011)}]{Kuroyanagi:2011iw}
\bibinfo{author}{\bibfnamefont{S.}~\bibnamefont{Kuroyanagi}} \bibnamefont{and}
  \bibinfo{author}{\bibfnamefont{T.}~\bibnamefont{Takahashi}},
  \bibinfo{journal}{JCAP} \textbf{\bibinfo{volume}{1110}}, \bibinfo{pages}{006}
  (\bibinfo{year}{2011}), \eprint{1106.3437}.

\bibitem[{\citenamefont{Leach and Liddle}(2001)}]{Leach:2000yw}
\bibinfo{author}{\bibfnamefont{S.~M.} \bibnamefont{Leach}} \bibnamefont{and}
  \bibinfo{author}{\bibfnamefont{A.~R.} \bibnamefont{Liddle}},
  \bibinfo{journal}{Phys. Rev.} \textbf{\bibinfo{volume}{D63}},
  \bibinfo{pages}{043508} (\bibinfo{year}{2001}), \eprint{astro-ph/0010082}.

\bibitem[{\citenamefont{Peiris and Easther}(2006)}]{Peiris:2006sj}
\bibinfo{author}{\bibfnamefont{H.}~\bibnamefont{Peiris}} \bibnamefont{and}
  \bibinfo{author}{\bibfnamefont{R.}~\bibnamefont{Easther}},
  \bibinfo{journal}{JCAP} \textbf{\bibinfo{volume}{0610}}, \bibinfo{pages}{017}
  (\bibinfo{year}{2006}), \eprint{astro-ph/0609003}.

\bibitem[{\citenamefont{Bean et~al.}(2008)\citenamefont{Bean, Chen, Peiris, and
  Xu}}]{Bean:2007eh}
\bibinfo{author}{\bibfnamefont{R.}~\bibnamefont{Bean}},
  \bibinfo{author}{\bibfnamefont{X.}~\bibnamefont{Chen}},
  \bibinfo{author}{\bibfnamefont{H.~V.} \bibnamefont{Peiris}},
  \bibnamefont{and} \bibinfo{author}{\bibfnamefont{J.}~\bibnamefont{Xu}},
  \bibinfo{journal}{Phys. Rev.} \textbf{\bibinfo{volume}{D77}},
  \bibinfo{pages}{023527} (\bibinfo{year}{2008}), \eprint{0710.1812}.

\bibitem[{\citenamefont{Lorenz et~al.}(2008)\citenamefont{Lorenz, Martin, and
  Ringeval}}]{Lorenz:2008je}
\bibinfo{author}{\bibfnamefont{L.}~\bibnamefont{Lorenz}},
  \bibinfo{author}{\bibfnamefont{J.}~\bibnamefont{Martin}}, \bibnamefont{and}
  \bibinfo{author}{\bibfnamefont{C.}~\bibnamefont{Ringeval}},
  \bibinfo{journal}{Phys. Rev.} \textbf{\bibinfo{volume}{D78}},
  \bibinfo{pages}{063543} (\bibinfo{year}{2008}), \eprint{0807.2414}.

\bibitem[{\citenamefont{Finelli et~al.}(2009)\citenamefont{Finelli, Hamann,
  Leach, and Lesgourgues}}]{Finelli:2009bs}
\bibinfo{author}{\bibfnamefont{F.}~\bibnamefont{Finelli}},
  \bibinfo{author}{\bibfnamefont{J.}~\bibnamefont{Hamann}},
  \bibinfo{author}{\bibfnamefont{S.~M.} \bibnamefont{Leach}}, \bibnamefont{and}
  \bibinfo{author}{\bibfnamefont{J.}~\bibnamefont{Lesgourgues}}
  (\bibinfo{year}{2009}), \eprint{0912.0522}.

\bibitem[{\citenamefont{Mortonson et~al.}(2011)\citenamefont{Mortonson, Peiris,
  and Easther}}]{Mortonson:2010er}
\bibinfo{author}{\bibfnamefont{M.~J.} \bibnamefont{Mortonson}},
  \bibinfo{author}{\bibfnamefont{H.~V.} \bibnamefont{Peiris}},
  \bibnamefont{and} \bibinfo{author}{\bibfnamefont{R.}~\bibnamefont{Easther}},
  \bibinfo{journal}{Phys. Rev.} \textbf{\bibinfo{volume}{D83}},
  \bibinfo{pages}{043505} (\bibinfo{year}{2011}), \eprint{1007.4205}.

\bibitem[{\citenamefont{Martin et~al.}(2011)\citenamefont{Martin, Ringeval, and
  Trotta}}]{Martin:2010hh}
\bibinfo{author}{\bibfnamefont{J.}~\bibnamefont{Martin}},
  \bibinfo{author}{\bibfnamefont{C.}~\bibnamefont{Ringeval}}, \bibnamefont{and}
  \bibinfo{author}{\bibfnamefont{R.}~\bibnamefont{Trotta}},
  \bibinfo{journal}{Phys.Rev.} \textbf{\bibinfo{volume}{D83}},
  \bibinfo{pages}{063524} (\bibinfo{year}{2011}), \eprint{1009.4157}.

\bibitem[{\citenamefont{Larson et~al.}(2011)}]{Larson:2010gs}
\bibinfo{author}{\bibfnamefont{D.}~\bibnamefont{Larson}} \bibnamefont{et~al.},
  \bibinfo{journal}{Astrophys. J. Suppl.} \textbf{\bibinfo{volume}{192}},
  \bibinfo{pages}{16} (\bibinfo{year}{2011}), \eprint{1001.4635}.

\bibitem[{\citenamefont{Jarosik et~al.}(2011)}]{Jarosik:2010iu}
\bibinfo{author}{\bibfnamefont{N.}~\bibnamefont{Jarosik}} \bibnamefont{et~al.},
  \bibinfo{journal}{Astrophys. J. Suppl.} \textbf{\bibinfo{volume}{192}},
  \bibinfo{pages}{14} (\bibinfo{year}{2011}), \eprint{1001.4744}.

\bibitem[{\citenamefont{Riess et~al.}(2009)}]{Riess:2009pu}
\bibinfo{author}{\bibfnamefont{A.~G.} \bibnamefont{Riess}}
  \bibnamefont{et~al.}, \bibinfo{journal}{Astrophys. J.}
  \textbf{\bibinfo{volume}{699}}, \bibinfo{pages}{539} (\bibinfo{year}{2009}),
  \eprint{0905.0695}.

\bibitem[{\citenamefont{Maggiore}(2000)}]{Maggiore:1999vm}
\bibinfo{author}{\bibfnamefont{M.}~\bibnamefont{Maggiore}},
  \bibinfo{journal}{Phys.Rept.} \textbf{\bibinfo{volume}{331}},
  \bibinfo{pages}{283} (\bibinfo{year}{2000}), \eprint{gr-qc/9909001}.

\bibitem[{\citenamefont{Kuroyanagi
  et~al.}(2011{\natexlab{a}})\citenamefont{Kuroyanagi, Chiba, and
  Sugiyama}}]{Kuroyanagi:2010mm}
\bibinfo{author}{\bibfnamefont{S.}~\bibnamefont{Kuroyanagi}},
  \bibinfo{author}{\bibfnamefont{T.}~\bibnamefont{Chiba}}, \bibnamefont{and}
  \bibinfo{author}{\bibfnamefont{N.}~\bibnamefont{Sugiyama}},
  \bibinfo{journal}{Phys.Rev.} \textbf{\bibinfo{volume}{D83}},
  \bibinfo{pages}{043514} (\bibinfo{year}{2011}{\natexlab{a}}),
  \eprint{1010.5246}.

\bibitem[{\citenamefont{Turner et~al.}(1993)\citenamefont{Turner, White, and
  Lidsey}}]{Turner:1993vb}
\bibinfo{author}{\bibfnamefont{M.~S.} \bibnamefont{Turner}},
  \bibinfo{author}{\bibfnamefont{M.~J.} \bibnamefont{White}}, \bibnamefont{and}
  \bibinfo{author}{\bibfnamefont{J.~E.} \bibnamefont{Lidsey}},
  \bibinfo{journal}{Phys.Rev.} \textbf{\bibinfo{volume}{D48}},
  \bibinfo{pages}{4613} (\bibinfo{year}{1993}), \eprint{astro-ph/9306029}.

\bibitem[{\citenamefont{Turner}(1983)}]{Turner:1983he}
\bibinfo{author}{\bibfnamefont{M.~S.} \bibnamefont{Turner}},
  \bibinfo{journal}{Phys. Rev.} \textbf{\bibinfo{volume}{D28}},
  \bibinfo{pages}{1243} (\bibinfo{year}{1983}).

\bibitem[{\citenamefont{Zaldarriaga et~al.}(1997)\citenamefont{Zaldarriaga,
  Spergel, and Seljak}}]{Zaldarriaga:1997ch}
\bibinfo{author}{\bibfnamefont{M.}~\bibnamefont{Zaldarriaga}},
  \bibinfo{author}{\bibfnamefont{D.~N.} \bibnamefont{Spergel}},
  \bibnamefont{and} \bibinfo{author}{\bibfnamefont{U.}~\bibnamefont{Seljak}},
  \bibinfo{journal}{Astrophys.J.} \textbf{\bibinfo{volume}{488}},
  \bibinfo{pages}{1} (\bibinfo{year}{1997}), \eprint{astro-ph/9702157}.

\bibitem[{\citenamefont{Seto}(2006)}]{Seto:2005qy}
\bibinfo{author}{\bibfnamefont{N.}~\bibnamefont{Seto}},
  \bibinfo{journal}{Phys.Rev.} \textbf{\bibinfo{volume}{D73}},
  \bibinfo{pages}{063001} (\bibinfo{year}{2006}), \eprint{gr-qc/0510067}.

\bibitem[{\citenamefont{Collaboration}(2006)}]{:2006uk}
\bibinfo{author}{\bibfnamefont{T.~P.} \bibnamefont{Collaboration}}
  (\bibinfo{collaboration}{Planck}) (\bibinfo{year}{2006}),
  \eprint{astro-ph/0604069}.

\bibitem[{\citenamefont{Baumann et~al.}(2009)}]{Baumann:2008aq}
\bibinfo{author}{\bibfnamefont{D.}~\bibnamefont{Baumann}} \bibnamefont{et~al.}
  (\bibinfo{collaboration}{CMBPol Study Team}), \bibinfo{journal}{AIP
  Conf.Proc.} \textbf{\bibinfo{volume}{1141}}, \bibinfo{pages}{10}
  (\bibinfo{year}{2009}), \eprint{0811.3919}.

\bibitem[{\citenamefont{{Bond} et~al.}(1997)\citenamefont{{Bond}, {Efstathiou},
  and {Tegmark}}}]{1997MNRAS.291L..33B}
\bibinfo{author}{\bibfnamefont{J.~R.} \bibnamefont{{Bond}}},
  \bibinfo{author}{\bibfnamefont{G.}~\bibnamefont{{Efstathiou}}},
  \bibnamefont{and}
  \bibinfo{author}{\bibfnamefont{M.}~\bibnamefont{{Tegmark}}},
  \bibinfo{journal}{Mon. Not. R. Astron. Soc.} \textbf{\bibinfo{volume}{291}},
  \bibinfo{pages}{L33} (\bibinfo{year}{1997}), \eprint{arXiv:astro-ph/9702100}.

\bibitem[{\citenamefont{Corbin and Cornish}(2006)}]{Corbin:2005ny}
\bibinfo{author}{\bibfnamefont{V.}~\bibnamefont{Corbin}} \bibnamefont{and}
  \bibinfo{author}{\bibfnamefont{N.~J.} \bibnamefont{Cornish}},
  \bibinfo{journal}{Class.Quant.Grav.} \textbf{\bibinfo{volume}{23}},
  \bibinfo{pages}{2435} (\bibinfo{year}{2006}), \eprint{gr-qc/0512039}.

\bibitem[{\citenamefont{Cutler and Harms}(2006)}]{Cutler:2005qq}
\bibinfo{author}{\bibfnamefont{C.}~\bibnamefont{Cutler}} \bibnamefont{and}
  \bibinfo{author}{\bibfnamefont{J.}~\bibnamefont{Harms}},
  \bibinfo{journal}{Phys. Rev.} \textbf{\bibinfo{volume}{D73}},
  \bibinfo{pages}{042001} (\bibinfo{year}{2006}), \eprint{gr-qc/0511092}.

\bibitem[{\citenamefont{Harms et~al.}(2008)\citenamefont{Harms, Mahrdt, Otto,
  and Priess}}]{Harms:2008xv}
\bibinfo{author}{\bibfnamefont{J.}~\bibnamefont{Harms}},
  \bibinfo{author}{\bibfnamefont{C.}~\bibnamefont{Mahrdt}},
  \bibinfo{author}{\bibfnamefont{M.}~\bibnamefont{Otto}}, \bibnamefont{and}
  \bibinfo{author}{\bibfnamefont{M.}~\bibnamefont{Priess}},
  \bibinfo{journal}{Phys. Rev.} \textbf{\bibinfo{volume}{D77}},
  \bibinfo{pages}{123010} (\bibinfo{year}{2008}), \eprint{0803.0226}.

\bibitem[{\citenamefont{Yagi and Seto}(2011)}]{Yagi:2011wg}
\bibinfo{author}{\bibfnamefont{K.}~\bibnamefont{Yagi}} \bibnamefont{and}
  \bibinfo{author}{\bibfnamefont{N.}~\bibnamefont{Seto}},
  \bibinfo{journal}{Phys. Rev.} \textbf{\bibinfo{volume}{D83}},
  \bibinfo{pages}{044011} (\bibinfo{year}{2011}), \eprint{1101.3940}.

\bibitem[{\citenamefont{Kosowsky
  et~al.}(1992{\natexlab{a}})\citenamefont{Kosowsky, Turner, and
  Watkins}}]{Kosowsky:1992rz}
\bibinfo{author}{\bibfnamefont{A.}~\bibnamefont{Kosowsky}},
  \bibinfo{author}{\bibfnamefont{M.~S.} \bibnamefont{Turner}},
  \bibnamefont{and} \bibinfo{author}{\bibfnamefont{R.}~\bibnamefont{Watkins}},
  \bibinfo{journal}{Phys.Rev.Lett.} \textbf{\bibinfo{volume}{69}},
  \bibinfo{pages}{2026} (\bibinfo{year}{1992}{\natexlab{a}}).

\bibitem[{\citenamefont{Kosowsky
  et~al.}(1992{\natexlab{b}})\citenamefont{Kosowsky, Turner, and
  Watkins}}]{Kosowsky:1991ua}
\bibinfo{author}{\bibfnamefont{A.}~\bibnamefont{Kosowsky}},
  \bibinfo{author}{\bibfnamefont{M.~S.} \bibnamefont{Turner}},
  \bibnamefont{and} \bibinfo{author}{\bibfnamefont{R.}~\bibnamefont{Watkins}},
  \bibinfo{journal}{Phys.Rev.} \textbf{\bibinfo{volume}{D45}},
  \bibinfo{pages}{4514} (\bibinfo{year}{1992}{\natexlab{b}}).

\bibitem[{\citenamefont{Kamionkowski et~al.}(1994)\citenamefont{Kamionkowski,
  Kosowsky, and Turner}}]{Kamionkowski:1993fg}
\bibinfo{author}{\bibfnamefont{M.}~\bibnamefont{Kamionkowski}},
  \bibinfo{author}{\bibfnamefont{A.}~\bibnamefont{Kosowsky}}, \bibnamefont{and}
  \bibinfo{author}{\bibfnamefont{M.~S.} \bibnamefont{Turner}},
  \bibinfo{journal}{Phys.Rev.} \textbf{\bibinfo{volume}{D49}},
  \bibinfo{pages}{2837} (\bibinfo{year}{1994}), \eprint{astro-ph/9310044}.

\bibitem[{\citenamefont{Caprini et~al.}(2008)\citenamefont{Caprini, Durrer, and
  Servant}}]{Caprini:2007xq}
\bibinfo{author}{\bibfnamefont{C.}~\bibnamefont{Caprini}},
  \bibinfo{author}{\bibfnamefont{R.}~\bibnamefont{Durrer}}, \bibnamefont{and}
  \bibinfo{author}{\bibfnamefont{G.}~\bibnamefont{Servant}},
  \bibinfo{journal}{Phys.Rev.} \textbf{\bibinfo{volume}{D77}},
  \bibinfo{pages}{124015} (\bibinfo{year}{2008}), \eprint{0711.2593}.

\bibitem[{\citenamefont{Caprini et~al.}(2009)\citenamefont{Caprini, Durrer, and
  Servant}}]{Caprini:2009yp}
\bibinfo{author}{\bibfnamefont{C.}~\bibnamefont{Caprini}},
  \bibinfo{author}{\bibfnamefont{R.}~\bibnamefont{Durrer}}, \bibnamefont{and}
  \bibinfo{author}{\bibfnamefont{G.}~\bibnamefont{Servant}},
  \bibinfo{journal}{JCAP} \textbf{\bibinfo{volume}{0912}}, \bibinfo{pages}{024}
  (\bibinfo{year}{2009}), \eprint{0909.0622}.

\bibitem[{\citenamefont{Khlebnikov and Tkachev}(1997)}]{Khlebnikov:1997di}
\bibinfo{author}{\bibfnamefont{S.}~\bibnamefont{Khlebnikov}} \bibnamefont{and}
  \bibinfo{author}{\bibfnamefont{I.}~\bibnamefont{Tkachev}},
  \bibinfo{journal}{Phys.Rev.} \textbf{\bibinfo{volume}{D56}},
  \bibinfo{pages}{653} (\bibinfo{year}{1997}), \eprint{hep-ph/9701423}.

\bibitem[{\citenamefont{Easther and Lim}(2006)}]{Easther:2006gt}
\bibinfo{author}{\bibfnamefont{R.}~\bibnamefont{Easther}} \bibnamefont{and}
  \bibinfo{author}{\bibfnamefont{E.~A.} \bibnamefont{Lim}},
  \bibinfo{journal}{JCAP} \textbf{\bibinfo{volume}{0604}}, \bibinfo{pages}{010}
  (\bibinfo{year}{2006}), \eprint{astro-ph/0601617}.

\bibitem[{\citenamefont{Garcia-Bellido and
  Figueroa}(2007)}]{GarciaBellido:2007dg}
\bibinfo{author}{\bibfnamefont{J.}~\bibnamefont{Garcia-Bellido}}
  \bibnamefont{and} \bibinfo{author}{\bibfnamefont{D.~G.}
  \bibnamefont{Figueroa}}, \bibinfo{journal}{Phys.Rev.Lett.}
  \textbf{\bibinfo{volume}{98}}, \bibinfo{pages}{061302}
  (\bibinfo{year}{2007}), \eprint{astro-ph/0701014}.

\bibitem[{\citenamefont{Garcia-Bellido
  et~al.}(2008)\citenamefont{Garcia-Bellido, Figueroa, and
  Sastre}}]{GarciaBellido:2007af}
\bibinfo{author}{\bibfnamefont{J.}~\bibnamefont{Garcia-Bellido}},
  \bibinfo{author}{\bibfnamefont{D.~G.} \bibnamefont{Figueroa}},
  \bibnamefont{and} \bibinfo{author}{\bibfnamefont{A.}~\bibnamefont{Sastre}},
  \bibinfo{journal}{Phys.Rev.} \textbf{\bibinfo{volume}{D77}},
  \bibinfo{pages}{043517} (\bibinfo{year}{2008}), \eprint{0707.0839}.

\bibitem[{\citenamefont{Dufaux et~al.}(2007)\citenamefont{Dufaux, Bergman,
  Felder, Kofman, and Uzan}}]{Dufaux:2007pt}
\bibinfo{author}{\bibfnamefont{J.~F.} \bibnamefont{Dufaux}},
  \bibinfo{author}{\bibfnamefont{A.}~\bibnamefont{Bergman}},
  \bibinfo{author}{\bibfnamefont{G.~N.} \bibnamefont{Felder}},
  \bibinfo{author}{\bibfnamefont{L.}~\bibnamefont{Kofman}}, \bibnamefont{and}
  \bibinfo{author}{\bibfnamefont{J.-P.} \bibnamefont{Uzan}},
  \bibinfo{journal}{Phys.Rev.} \textbf{\bibinfo{volume}{D76}},
  \bibinfo{pages}{123517} (\bibinfo{year}{2007}), \eprint{0707.0875}.

\bibitem[{\citenamefont{Barnaby et~al.}(2012)\citenamefont{Barnaby, Moxon,
  Namba, Peloso, Shiu et~al.}}]{Barnaby:2012xt}
\bibinfo{author}{\bibfnamefont{N.}~\bibnamefont{Barnaby}},
  \bibinfo{author}{\bibfnamefont{J.}~\bibnamefont{Moxon}},
  \bibinfo{author}{\bibfnamefont{R.}~\bibnamefont{Namba}},
  \bibinfo{author}{\bibfnamefont{M.}~\bibnamefont{Peloso}},
  \bibinfo{author}{\bibfnamefont{G.}~\bibnamefont{Shiu}}, \bibnamefont{et~al.},
  \bibinfo{journal}{Phys.Rev.} \textbf{\bibinfo{volume}{D86}},
  \bibinfo{pages}{103508} (\bibinfo{year}{2012}), \eprint{1206.6117}.

\bibitem[{\citenamefont{Vilenkin}(1981)}]{Vilenkin:1981bx}
\bibinfo{author}{\bibfnamefont{A.}~\bibnamefont{Vilenkin}},
  \bibinfo{journal}{Phys.Lett.} \textbf{\bibinfo{volume}{B107}},
  \bibinfo{pages}{47} (\bibinfo{year}{1981}).

\bibitem[{\citenamefont{Hogan and Rees}(1984)}]{Hogan:1984is}
\bibinfo{author}{\bibfnamefont{C.}~\bibnamefont{Hogan}} \bibnamefont{and}
  \bibinfo{author}{\bibfnamefont{M.}~\bibnamefont{Rees}},
  \bibinfo{journal}{Nature} \textbf{\bibinfo{volume}{311}},
  \bibinfo{pages}{109} (\bibinfo{year}{1984}).

\bibitem[{\citenamefont{Vachaspati and Vilenkin}(1985)}]{Vachaspati:1984gt}
\bibinfo{author}{\bibfnamefont{T.}~\bibnamefont{Vachaspati}} \bibnamefont{and}
  \bibinfo{author}{\bibfnamefont{A.}~\bibnamefont{Vilenkin}},
  \bibinfo{journal}{Phys.Rev.} \textbf{\bibinfo{volume}{D31}},
  \bibinfo{pages}{3052} (\bibinfo{year}{1985}).

\bibitem[{\citenamefont{Damour and Vilenkin}(2001)}]{Damour:2001bk}
\bibinfo{author}{\bibfnamefont{T.}~\bibnamefont{Damour}} \bibnamefont{and}
  \bibinfo{author}{\bibfnamefont{A.}~\bibnamefont{Vilenkin}},
  \bibinfo{journal}{Phys.Rev.} \textbf{\bibinfo{volume}{D64}},
  \bibinfo{pages}{064008} (\bibinfo{year}{2001}), \eprint{gr-qc/0104026}.

\bibitem[{\citenamefont{Olmez et~al.}(2010)\citenamefont{Olmez, Mandic, and
  Siemens}}]{Olmez:2010bi}
\bibinfo{author}{\bibfnamefont{S.}~\bibnamefont{Olmez}},
  \bibinfo{author}{\bibfnamefont{V.}~\bibnamefont{Mandic}}, \bibnamefont{and}
  \bibinfo{author}{\bibfnamefont{X.}~\bibnamefont{Siemens}},
  \bibinfo{journal}{Phys.Rev.} \textbf{\bibinfo{volume}{D81}},
  \bibinfo{pages}{104028} (\bibinfo{year}{2010}), \eprint{1004.0890}.

\bibitem[{\citenamefont{Kudoh et~al.}(2006)\citenamefont{Kudoh, Taruya,
  Hiramatsu, and Himemoto}}]{Kudoh:2005as}
\bibinfo{author}{\bibfnamefont{H.}~\bibnamefont{Kudoh}},
  \bibinfo{author}{\bibfnamefont{A.}~\bibnamefont{Taruya}},
  \bibinfo{author}{\bibfnamefont{T.}~\bibnamefont{Hiramatsu}},
  \bibnamefont{and} \bibinfo{author}{\bibfnamefont{Y.}~\bibnamefont{Himemoto}},
  \bibinfo{journal}{Phys. Rev.} \textbf{\bibinfo{volume}{D73}},
  \bibinfo{pages}{064006} (\bibinfo{year}{2006}), \eprint{gr-qc/0511145}.

\bibitem[{\citenamefont{Kuroyanagi
  et~al.}(2011{\natexlab{b}})\citenamefont{Kuroyanagi, Nakayama, and
  Saito}}]{Kuroyanagi:2011fy}
\bibinfo{author}{\bibfnamefont{S.}~\bibnamefont{Kuroyanagi}},
  \bibinfo{author}{\bibfnamefont{K.}~\bibnamefont{Nakayama}}, \bibnamefont{and}
  \bibinfo{author}{\bibfnamefont{S.}~\bibnamefont{Saito}},
  \bibinfo{journal}{Phys.Rev.} \textbf{\bibinfo{volume}{D84}},
  \bibinfo{pages}{123513} (\bibinfo{year}{2011}{\natexlab{b}}),
  \bibinfo{note}{21 pages, 8 figures}, \eprint{1110.4169}.

\end{thebibliography}

\end{document}